\documentclass[pdflatex,sn-mathphys-num]{sn-jnl}

\usepackage{graphicx}%
\usepackage{multirow}%
\usepackage{amsmath,amssymb,amsfonts}%
\usepackage{amsthm}%
\usepackage{mathrsfs}%
\usepackage[title]{appendix}%
\usepackage{xcolor}%
\usepackage{textcomp}%
\usepackage{manyfoot}%
\usepackage{booktabs}%
\usepackage{algorithm}%
\usepackage{algorithmicx}%
\usepackage{algpseudocode}%
\usepackage{listings}%

 \usepackage{bbm}

\newcommand{\be}{\begin{equation}}
\newcommand{\ee}{\end{equation}}

\newcommand{\beqa}{\begin{eqnarray}}
\newcommand{\eeqa}{\end{eqnarray}}
\newcommand{\nn}{\nonumber}

\def\CD {{\cal D}}
\def\CE {{\cal E}}

\def\CI {{\cal I}}

\def\CR {{\cal R}}
\def\CS {{\cal S}}

\theoremstyle{thmstyleone}%
\theoremstyle{thmstyletwo}%

\theoremstyle{thmstylethree}%

\begin{document}

\title[Absorption of Gravitons and  Photon Luminosities]{Absorption of Gravitons and  Photon Luminosities of Interstellar/Intergalactic Hydrogen Atoms

}


\author[1,3]{\fnm{George} \sur{Savvidy}}\email{savvidy@inp.demokritos.gr}

\author[2]{\fnm{Pavlos} \sur{Savvidis}}\email{p.savvidis@westlake.edu.cn}


\affil[1]{\orgdiv{Institute of Nuclear and Particle Physics}, \orgname{NCSR "Demokritos"}, \orgaddress{\street{Ag. Paraskevi}, \city{Athens}, \postcode{15342},  \country{Greece}}}

\affil[2]{\orgdiv{School of Science, Quantum Optoelectronics Lab}, \orgname{Westlake University}, \orgaddress{\street{Street}, \city{Hangzhou}, \postcode{310030},  \country{ China}}}

\affil[3]{\orgdiv{School of Physics and Astronomy}, \orgname{ Sun Yat-Sen University}, \orgaddress{\street{Street}, \city{Zhuhai}, \postcode{519082}, \country{China}}}


\abstract{ We compute the emission  and absorption rates of gravitons by hydrogen atoms.  The absorption rate of gravitons is  proportional to the number of hydrogen atoms and to the graviton luminosity. The number of hydrogen atoms in  interstellar/intergalactic media can be  large  while  the graviton luminosity of Sun, or a typical star, is  in  the eV to keV range that well overlaps with the resonance frequencies  of the hydrogen atoms.   Currently, the observational knowledge of the graviton luminosity  in different regions of the Universe  is limited, and here we will consider its value as unknown quantity that should be determined  by independent astrophysical observations.  We suggest measuring  the ratio of the photon luminosities  from  interstellar/intergalactic hydrogen atoms that can provide information concerning  the  luminosity  of gravitons in different regions of space.   This ratio  maximally exposes the fundamental physical nature of photons and gravitons – their helicity, which is h=1 for the  photons and h=2  for the gravitons. The ratio is sensitive to any sources of gravitational radiation and provide a possible method for measuring their radiation intensities.  }




\maketitle

\section{\it  Introduction}

We suggest a possible method for experimental detection of intensity of the gravitational radiation  by  measuring the ratio of photon luminosities from interstellar and intergalactic hydrogen atoms that are widespread  in large clouds  throughout the vast interstellar and intergalactic space.  The absorption rate of gravitons by the hydrogen atoms is proportional to number of atoms in a cloud and to the graviton luminosity.     Currently, the observational knowledge of the background energy density of gravitons in different regions of the Universe is limited, and here we will consider its value as an unknown quantity that should be determined  by independent astrophysical observations. This ratio of photon luminosities may allow to measure the intensity of the gravitational radiation.

In a series of publications  \cite{Weinberg:1964kqu, Weinberg:1964ew, Weinberg:1965nx, Weinberg:1965rz,Gould}  Weinberg and Gould estimated the power of  thermal radiation of gravitons generated by the Sun (or a typical star) through the scattering of electrons and protons in a completely ionised hydrogen plasma at the core of the Sun. The thermal graviton energies form a continuum spectrum and are in the $eV$  to $keV$ region.  A recent comprehensive  analysis of the gravitational emission of the Sun defines partial contributions from the photon-production,  $ee$ bremsstrahlung,  $ep$ bremsstrahlung,   $eHe$ bremsstrahlung and the emission induced by  the   hydrodynamical fluctuations  \cite{Garcia-Cely:2024ujr}.  This energy region very well matches with the absorption spectrum of the interstellar hydrogen atoms. 

The   spontaneous radiation rates of gravitons from the excited states of hydrogen atoms are tiny;  instead,  the  absorption rates of gravitons by hydrogen atoms is  numerically larger  than the spontaneous radiation rates.  In addition, the absorption rate is amplified  proportionally to the number of atoms in a hydrogen cloud and to the energy density of the graviton radiation.  These quantities can be large due to the fact that in the interstellar and intergalactic media  there are giant halos  containing a vast amount of hydrogen  atoms \cite{Dyson2020,Peebles1969ApJ}  and due to the  alternative  sources of gravitational radiation \cite{Ghiglieri_2015, Ringwald_2021, Grishchuk_1989, PhysRevLett.110.071105, Dyson:2013hbl}.

The problem  of measuring the graviton radiation was discussed in the literature by Dyson \cite{Dyson:2013hbl} and other authors \cite{ Tobar:2023ksi, Rothman:2006fp, Boughn:2006st, Marletto:2017kzi, Carney:2024dsj, Manikandan:2025qgv}. These investigations are  mostly concerned with a construction of sensitive ground-based  detectors that would be able to capture gravitational waves and gravitons from the in-falling gravitational waves. Here, instead, we suggest measuring the excess in the ratio $\CR_{\gamma}$ of the photon luminosities from the wast amount  of hydrogen atoms scattered in the interstellar and intergalactic space.

The article is organised as follows. In the second and the third sections we calculate the spontaneous emission and absorption rates of gravitons by the hydrogen atoms in  the $\vert n \geq 3 \rangle$  and   $\vert 1s \rangle$ states.      The discussion of thermal graviton luminosity of a typical star  and of a typical galaxy is presented in the fourth section. 

In the  fifth and sixth sections we define the ratio $\CR_{\gamma}$ that measures the  spontaneously radiated photons from the states $n\geq 3$. The ratio $\CR_{\gamma}$  is defined in a way  that maximally exposes the fundamental difference in the nature of photons and gravitons - their helicity, which is $h=1$ for the  photons and $h=2$  for the gravitons \cite{Savvidy:2025rqt}.   Because of this difference in their helicities   the spontaneous radiation luminosity $L^{\gamma}_{21}$ of the Lyman-$Ly_{\alpha}$ photons is induced exclusively  by the absorption of photons, while  the radiation of the Lyman-$Ly_{\beta}$ and Balmer-$H_{\alpha}$ photons from the $n=3$ state, the luminosity $L^{\gamma}_{31} + L^{\gamma}_{32}$, has the contributions from the absorption of both quanta: the photons  and the gravitons. This difference in the origin of photons  that are radiated from the  $n=2$ and $n=3$ states is used in the definition of the ratio $\CR_{\gamma}$ to measure the  photon luminosity from the $n=3$ state compared with the $n=2$ state. This ratio can be computed and has the following form
\be\label{mainratio}
\CR_{\gamma} = \frac{L^{\gamma}_{31} + L^{\gamma}_{32}}{L^{\gamma}_{21} }  = C_{31}  ~\frac{ U_{\gamma}(\omega_{31})}{U_{\gamma}(\omega_{21})} + C_{32} ~\frac{ U_{\gamma}(\omega_{32})}{U_{\gamma}(\omega_{21})},
\ee
where $U_{\gamma}(\omega_{nm})$ is the photon spectral density in a  given region and $C_{ij}$ are numerical coefficients. The  excess $\delta \CR = \CR_{obs} - \CR_{\gamma} $   would indicate the additional absorption to the $n=3$ state that may be attributed to the  gravitons, the massless particles of helicity $h=2$. This ratio is sensitive to any source of gravitons, and it is defined without any reference to the origin and intensity of a source.

 \section{\it  Graviton emission rates }
 
The energy density of the classical quadrupole gravitational radiation  has the following form \cite{Einstein2,Landau1975}:
\be\label{Einstein2}
\bar{\frac{d \CE}{d t}} = \frac{G}{45 c^5} \dddot{D}_{ij}\dddot{D}_{ij},
\ee
where the quadrupole momentum tensor  is
\be
D_{ij}= \int \rho (3 x_i x_j -\delta_{ij} \vec{x}^2) d V
\ee
and $\rho$ is the density of mass.     Let us  define the quantum-mechanical transition amplitude by the generalisation of  the expression (\ref{Einstein2}).   The quantum-mechanical  quadrupole  transition rate $\CS$ of an electron from  the state  $m$ to the state $n$ with the radiation of graviton will be: 
 \be\label{QMEinstein}
(Transition~ Rate~  m \rightarrow n)_g =  \frac{1}{\hbar \omega_{mn}}\bar{\frac{d \CE}{d t}} = \frac{G}{45 c^5 \hbar \omega_{mn} } ~ \vert \langle n \vert \dddot{D}_{ij} \vert m \rangle  \vert^2 ,
\ee
 and taking into account the time dependence of the wave functions $e^{i \omega_{mn} t}$ for the $\CS$
 we will get\footnote{A field-theoretical derivation of the formula (\ref{QMEinstein2}) can be found in    \cite{Weinberg:1972kfs} on page 286, formula (10.8.6) and in \cite{Gould}. The quantum mechanical effects such as fine structure, the Lamb shift, and hyperfine structure are not considered in this article.  This will be part of a future investigation. }
 \be\label{QMEinstein2}
\CS_g(m \rightarrow n) =  \frac{2 G \omega^5_{mn}}{45 \hbar c^5} ~ \vert \langle n \vert  D_{ij} \vert m \rangle  \vert^2,
\ee
 where $ \hbar \omega_{mn} = E^{(m)} - E^{(n)}  $.   For the hydrogen atom the matrix elements  $ \langle 1s \vert D_{ij} \vert 1s \rangle =0$  vanish   and   the transition rate from the $\vert  2p \rangle $   to the $\vert  1s \rangle $ state also vanishes: $  \langle 2p  \vert D_{ij} \vert 1s \rangle = 0$.   There is no radiation of gravitons from $n=1$ and $n=2$ states. This is a consequence of the fact that a massless graviton  has the helicity $h=2 $  \cite{Savvidy:2025rqt}, and nonzero transitions can only appear  from the $\vert 3d \rangle$  state.  The first nonzero radiation of gravitons  appears from the $\vert 3d \rangle$ to the $\vert 1s \rangle$ state and the corresponding matrix elements  are defined as
\beqa\label{matrixelem}
 &&\langle 3d_{l_z} \vert D_{ij} \vert 1s \rangle = m \int   R_{32} Y_{2,l_z}  ~  (3 x_{i} x_{i} - \delta_{ij} r^2)   R_{10} Y_{0,0} ~r^2 dr \sin \theta d\theta d \phi= \CD_{l_z}(ij),~~~\nn\\
&& l_z =0,\pm1,\pm2~~~~~i,j=1,2,3, 
 \eeqa
 and are given in the Appendix A.  Using these matrix elements (\ref{matquanra}) for the quadrupole momentum one can obtain the spontaneous transition rate of gravitons from the $\vert 3d \rangle$ to the  $\vert1s \rangle$ state:
 \be\label{spontaneous2}
\CS_g(3d \rightarrow 1s)= \frac{6561}{8192}   \frac{G m^2}{\hbar c} \frac{a^4 \omega^5_{31} }{ c^4} . 
\ee
Substituting the value of the angular frequency $\omega_{31}$  
\be\label{frequan2}
\omega_{31} = \frac{4 m e^4 }{9 \hbar^3}
\ee
and the Bohr radius $a$ into the above formula, we obtain the total spontaneous transition rate $\CS$  \cite{Weinberg:1972kfs,Gould}\footnote{The total transition rate is a sum of partial transition rates over all final and initial states.} :
\be\label{spontaneous1}
\CS_g(3d \rightarrow 1s)=    \frac{1}{36}  \frac{G m^2}{  \hbar c } \Big(\frac{ e^2}{ \hbar c} \Big)^4 \Big( \frac{m e^4}{2  \hbar^3} \Big) . 
\ee
 The first term is the mass of the electron in units of the Planck mass $M^2_{Pl}= c \hbar / G $, the second term is the electromagnetic fine-structure constant $\alpha = e^2/ \hbar c$, and the last term is the Rydberg constant divided by Planck constant. It is a beautiful unification of gravity,  electrodynamics and quantum mechanics in one expression.    The numerical value of the spontaneous radiation rate of a graviton by a hydrogen atom is 
 \be\label{grtrans1}
 \CS_g(3d \rightarrow 1s) \approx 2.85 \times 10^{-39} \frac{1}{sec}.
 \ee 
 This rate is very small and seems undetectable \cite{Dyson:2013hbl}. The life time of the $3d$  state due to the graviton radiation will be
 \be
 \tau = \frac{1}{\CS_g(3d \rightarrow 1s)} \approx 3.51 \times 10^{38} sec.
 \ee
The next nonzero radiation of gravitons can appear from the $\vert 4d \rangle$ to the $\vert 1s \rangle$ state.
These matrix elements  are defined as in (\ref{matrixelem}).  For the quadrupole momentum we obtain spontaneous radiation  rate of gravitons from $\vert 4d \rangle$ to  $\vert1s \rangle$    as
 \be\label{spontaneous2}
\CS_g(4d \rightarrow 1s)= \frac{51539607552}{152587890625}   \frac{G m^2}{\hbar c} \frac{a^4 \omega^5_{41} }{ c^4} . 
\ee
Substituting the value of the Bohr radius and of the angular frequency  
\be\label{frequan2}
\omega_{41} = \frac{15 m e^4 }{32 \hbar^3}
\ee
we obtain spontaneous transition rate $\CS$:
\be\label{spontaneous2}
\CS_g(4d \rightarrow 1s)=    \frac{2^{10} 3^6}{5^{11}}  \frac{G m^2}{  \hbar c } \Big(\frac{ e^2}{ \hbar c} \Big)^4 \Big( \frac{m e^4}{2  \hbar^3} \Big). 
\ee
The numerical value of the spontaneous radiation rate is 
 \be\label{grtrans2}
 \CS_g(4d \rightarrow 1s) \approx 1.57 \times 10^{-39} \frac{1}{sec} .
 \ee 
 It is of the same order of magnitude as $ ( Rate~ 3d \rightarrow 1s)_{g} $ in (\ref{grtrans1}), and the life time of the $4d$  state due to the graviton radiation will be
 \be
 \tau = \frac{1}{\CS_g(4d \rightarrow 1s)} \approx 6.37 \times 10^{38} sec.
 \ee
 In the course of the cosmological expansion at the time of recombination the electrons are captured by protons, and these initial bound states appear at  $n \approx 350$. Even higher energy states with $n=500-1000$ were achieved  in the laboratory experiments \cite{Fischer_1994}. It seems therefore reasonable to consider the  interaction of gravitons with high excited states of the hydrogen atoms as well. The radiation rate (\ref{QMEinstein2}) for high excited states has the following form:
 \be\label{spontaneousn}
\CS_g(nd \rightarrow 1s) \approx        \frac{2^{2n+4} ~n^6   }{  15  (n+1)^{8+2n}   } \frac{  \Gamma[n+4]^2 }{  \Gamma[2n] }    \frac{G m^2}{  \hbar c }   \frac{a^4 \omega^5_{n1} }{ c^4}  ,
\ee 
where the square of the matrix element is\footnote{In the above approximation we use the asymptotic form of the $R_{n2}$ wave function at $r \gg a$  (see Appendix A). }
\beqa
\vert \langle nd  \vert D_{ij} \vert 1s \rangle\vert^2 &&= \sum_{l_z=0\pm1\pm2}\vert  m \int   R_{n2} Y_{2,l_z}  ~  (3 x_{i} x_{i} - \delta_{ij} r^2)   R_{10} Y_{0,0} ~r^2 dr \sin \theta d\theta d \phi \vert^2 \nn\\
&&\approx  \frac{  3 ~2^{2n+3} ~n^6   }{    (n+1)^{8+2n}   }     \frac{  \Gamma[n+4]^2 }{  \Gamma[2n] }  m^2 a^4 
\eeqa
and 
\be\label{frequ}
\omega_{n1}=  \frac{n^2-1}{n^2}  \frac{m e^4}{2 \hbar^3} .
\ee    
Substituting the value of the Bohr radius and of the angular frequency $\omega_{n1}$  we will get 
\be\label{spontaneousn1}
\CS_g(nd \rightarrow 1s)  \approx     \frac{2^{2n+1}   (n   -1)^5}{15 n^3 (n+1)^{2n+3}}    \frac{  \Gamma[n+4]^2 }{  \Gamma[2n+1] }   \frac{G m^2}{  \hbar c } \Big(\frac{ e^2}{ \hbar c} \Big)^4 \Big( \frac{m e^4}{2  \hbar^3} \Big). 
\ee 
The numerical value of the spontaneous radiation rates for the low lying levels $n$ are given below:   \beqa\label{grtransn}
 \CS_g(3d \rightarrow 1s) \approx 2.85 \times 10^{-39} \frac{1}{sec} ,\nn\\
\CS_g(4d \rightarrow 1s)   \approx 1.72  \times 10^{-40} \frac{1}{sec} ,\nn\\
 ...............
 \eeqa  
 These  transition rates are decreasing as $ \frac{2 \pi^{5/2}}{15 n^{2n +3/2}}$.  A faster decrease of transition probabilities  appears in (\ref{grtransn}) because of the approximation that we were using for the wave function $R_{n2}$ (\ref{append}).  This approximation for the wave function has lower values near $r \approx a$,  while the  exact values of $R_{n2}$ near  $r \approx a$ are  larger.

As it follows from the above consideration,  the   spontaneous radiation rate $\CS_g$ of gravitons from the excited states of the hydrogen atoms is tiny  (\ref{QMEinstein2}), (\ref{grtrans1}) and (\ref{grtrans2}).  Instead,  as we will show in the nest section, the   absorption rates $\CD_g$ of gravitons by hydrogen atoms  (\ref{fundabsorbrate}) are numerically larger  than the spontaneous radiation rates $\CS_g$.  In addition, the absorption rates are amplified  proportionally to the number of hydrogen atoms $N_{1s}$  in the interstellar and intergalactic clouds and to the spectral energy density $U_g(\omega)$ of the graviton radiation (\ref{absorp}).  These quantities can be large due to the fact that in the interstellar and intergalactic media  there are giant halos  containing a vast amount of hydrogen  atoms \cite{Dyson2020,Peebles1969ApJ}  and due to the alternative  sources of gravitational radiation \cite{Ghiglieri_2015, Ringwald_2021, Grishchuk_1989, PhysRevLett.110.071105, Dyson:2013hbl}. 

The giant halos of pristine, primordial hydrogen gas that surrounded galaxies in the early universe  at z = 5.5 - 13.4  were found by the James Webb Space Telescope\cite{2023A&A...677A..88B,  2025A&A...693A..60H,  Kashino_2023,heintz2023,  2024ApJ...962...24S}. The presents of copious amount of pristine HI gas was discovered  by measurement of  the  strong Ly$\alpha$ emission and  absorption line in a large spectroscopic sample of star-forming galaxies at $z = 5.5 - 13.4$ observed with JWST\cite{2023A&A...677A..88B,  2025A&A...693A..60H,  Kashino_2023,heintz2023,  2024ApJ...962...24S}.

\section{\it  Graviton absorption rates }

\begin{figure}
 \centering
\includegraphics[angle=0,width=9cm]{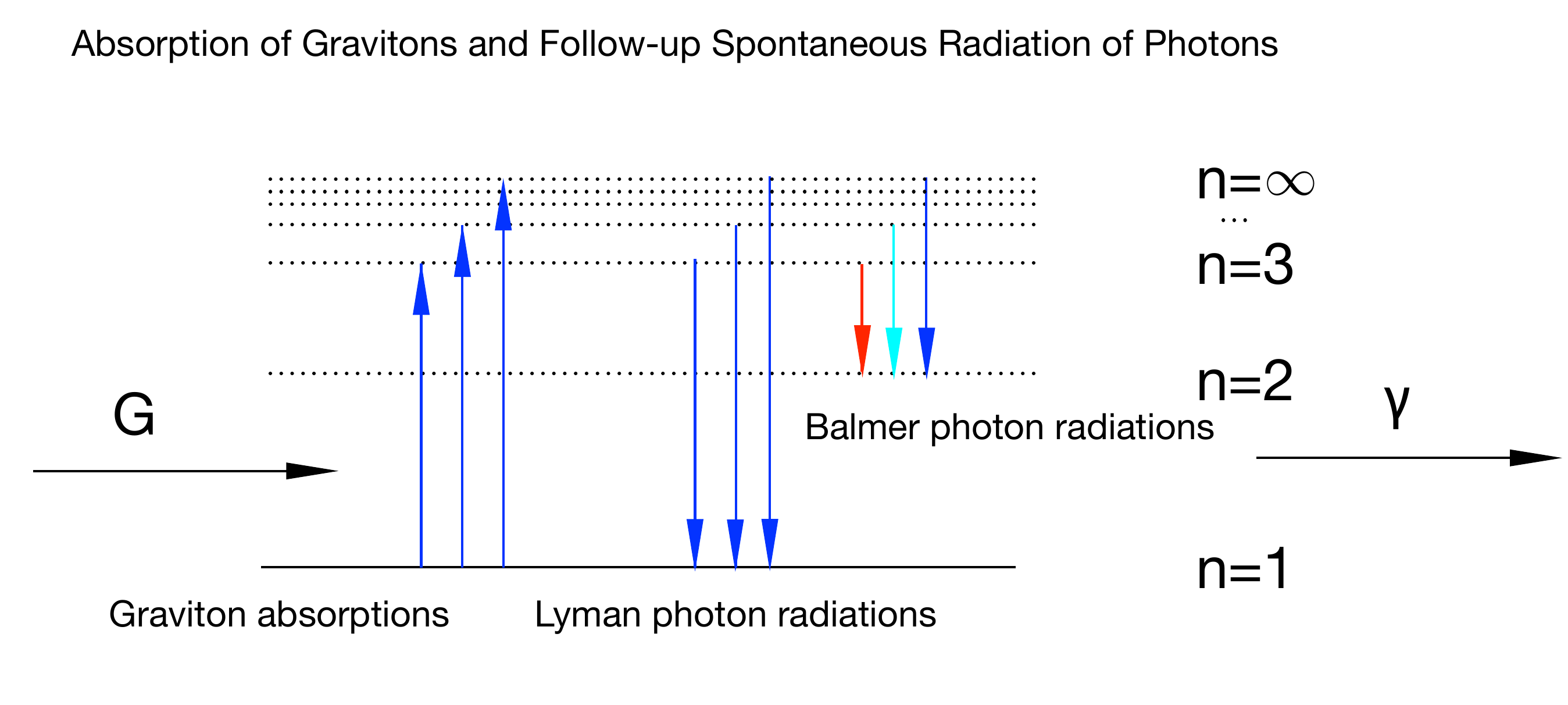} ~~
\centering
\caption{The figure demonstrates the absorption of gravitons $G$ and the follow-up  spontaneous radiation of photons $\gamma$.  There is no absorption of gravitons to the $n=2$ state (\ref{zerotrans}).  The graviton absorption wave lengths are in the UV region  $91.1~nm \leq \lambda_g \leq 102.5 ~nm$ and very well overlap  with the spectrum of the graviton radiation of stars $1.88~ nm <  \lambda^{\odot}_{g} <  1884~nm$ (\ref{lambdagrange}).   }
\label{fig1} 
\end{figure}

 In order to obtain the rate of graviton  absorption we will consider an extension of the Einstein derivation of absorption and stimulated radiation rates for photons in equilibrium with  thermal bath at temperature $k T$ \cite{Einstein:1917zz,Weinberg:1972kfs}. The rate of graviton absorption can be represented in  the following form  \cite{Dirac:1927dy,Fermi:1932xva, Fermi1954}:
 \be\label{absorp}
 (Absorption~rate~of~ n \rightarrow m)_{g} = \CD_g ~U_g(\omega) N(n),
 \ee
 where $U_g(\omega) d\omega$ is the energy density of the infalling gravitational radiation and $N(n)$ is the number of atoms in the state $n$. The total radiation transition rate of the excited atoms is a sum of spontaneous transitions $\CS_g$  (\ref{QMEinstein2}) and stimulated transitions $\CI$:
\be\label{equa1}
 (Total~radiation~rate~of~ m \rightarrow n)_g = \CS + \CI ~  U_g(\omega) N(m).
 \ee 
 The equation of the detailed balance at the equilibrium $(Rate~n \rightarrow m)_g = (Rate~m \rightarrow n )_g $ is
 \be\label{equa2}
 \CD_g~ U_g(\omega) N(n) = \CS  + \CI_g~ U_g(\omega) N(m). 
 \ee
 The energy density of the gravitational field follows the  Planck distribution due to the Bose statistics of the gas of gravitons:
 \be\label{equa3}
 U_g(\omega) = \frac{\hbar \omega^3}{\pi^2  c^3}\frac{1}{e^{\frac{\hbar \omega}{k T}    }-1},
 \ee
 and the ratio of atoms in the states $n$ and $m$ is given by the Boltzmann  distribution:
 \be\label{equa4}
 \frac{N(n)}{N(m)} = e^{-\frac{(E^{(m)} - E^{(n)}}{k T} }= e^{-\frac{\hbar \omega_{mn}}{k T}}.
 \ee
 Substituting (\ref{equa3}) and (\ref{equa4}) into (\ref{equa2}) we obtain the expression for absorption and stimulated emission rates in terms of the known spontaneous  transition rate $\CS$ (\ref{QMEinstein2}): 
 \be\label{fundabsor}
 \CD_g= \frac{\pi^2 c^3}{\hbar \omega^3} \CS_g,~~~~~\CI_g = \CD_g.   
 \ee
 Thus we obtained the main expression for the absorption rate of gravitons 
\be\label{fundabsorbrate}
 \CD_g= \frac{2 \pi^2  }{45  }  \frac{  G \omega^2_{mn}}{  \hbar^2 c^2} ~ \vert \langle n \vert  D_{ij} \vert m \rangle  \vert^2. 
 \ee
 As we discussed above the graviton helicity is $h=\pm 2$ \cite{Savvidy:2025rqt} and  there is no absorption of gravitons to the $n=2$ state because $  \langle 2  \vert D_{ij} \vert 1s \rangle = 0$:
 \be \label{zerotrans}
  \CD_g(1s \rightarrow 2)  =   0.
 \ee
 We can now obtain the expression for the absorption rates using the formulas (\ref{spontaneous1}), (\ref{spontaneous2}) and (\ref{spontaneousn1})  for $\CS$:
 \be 
  \CD_g(1s \rightarrow 3d)  =   \frac{6561 \pi^2}{8192}   \frac{ G m^2}{\hbar c  } \frac{a^4 \omega^2_{31} }{ \hbar c}.
 \ee
 Substituting the angular frequency $\omega_{31}$ from (\ref{frequan2}) and the Bohr radius $a$ we will obtain the absorption rate  of gravitons:
 \be
 \CD_g(1s \rightarrow 3d) =  \frac{81  }{512}   \frac{G \pi^2}{c^2}.  
\ee
The value of this graviton absorption rate by a hydrogen atom is 
 \be\label{onetothree}
 \CD_g(1s \rightarrow 3d) \approx 1.16 \times 10^{-28}~ \frac{cm}{gram}.
 \ee 
This rate is numerically large than the spontaneous radiation rate (\ref{grtrans1})    (see Fig.\ref{fig1},\ref{fig2},\ref{fig3},\ref{fig4}):
\beqa\label{absorategrav}
 (Rate~1s \rightarrow 3d)_{g} = \frac{81  }{512}   \frac{G \pi^2}{c^2} ~ N_{1s}~U_g(\omega_{31}).    
\eeqa
The number of  hydrogen atoms $N_{1s}$ in the $\vert 1s\rangle$ state in a container of $10^6$ grams of the hydrogen gas would contain  $N_{1s}= 10^{6} N_A$ atoms, where the Avogadro number is $N_A= 6.0221\times 10^{24}$, and if in addition there is a nonzero energy density of gravitons $U(\omega_{31}) d \omega$, then the absorption rate would  be 
\beqa\label{absorategravnum3}
(Rate~1s \rightarrow 3d)_{g} =\Big(\frac{U_g(\omega_{31})  }{gram/cm ~sec}  \Big) \Big(\frac{mass~of~H}{10^6 ~gram}\Big) \times 6.98 \times 10^{2}  sec^{-1}.
\eeqa
 One can also obtain the expression for the absorption  rate for the $\vert 1s \rangle$ to  $\vert 4d \rangle$ by using the formula  (\ref{spontaneous2})  for $\CS$:
 \be 
  \CD_g(1s \rightarrow 4d)  =       \frac{51539607552 \pi^2}{152587890625}   \frac{ G m^2}{\hbar c} \frac{a^4 \omega^2_{41} }{ \hbar c}.
 \ee
 Substituting the angular frequency $\omega_{41}$  (\ref{frequan2}) and the Bohr radius $a$ we will obtain our formula for the absorption rate  of gravitons\footnote{The coefficient in (\ref{14}) is $2^{24} 3^3/ 5^{14}$.}:
 \be\label{14}
 \CD_g(1s \rightarrow 4d) =  \frac{452984832 }{6103515625}   \frac{G  \pi^2}{c^2},
\ee
so that this graviton absorption rate by a hydrogen atom is 
 \be
 \CD_g(1s \rightarrow 4d) \approx 5.44 \times 10^{-29}~ \frac{cm}{gram}.
 \ee 
This rate is of the same order as the  $\vert 1s \rangle$ to  $\vert 3d \rangle$  absorption rate  (\ref{onetothree}), and we have 
 \beqa\label{absorategrav2}
 (Rate~1s \rightarrow 4d)_{g} =  \frac{  2^{24} 3^3 }{5^{14}}   \frac{G \pi^2}{c^2} ~ N_{1s}~U_g(\omega_{41}).    
\eeqa
For $10^6$ grams of the neutral hydrogen gas and with a nonzero energy density of gravitons $U_g(\omega_{41}) d \omega$   the absorption rate would  be 
\beqa\label{absorategravnum4}
(Rate~1s \rightarrow 4d)_{g} =\Big(\frac{U_g(\omega_{41})  }{gram/cm ~sec}  \Big) \Big(\frac{mass~of~H}{10^6 ~gram}\Big) \times 3.27 \times 10^{2}  sec^{-1}.
\eeqa
We can obtain the absorption rate to the high-energy levels by using the formula (\ref{spontaneousn})
\be 
  \CD_g(1s \rightarrow nd)   \approx        \frac{ \pi^2~2^{2n+4} n^6   }{  15  (n+1)^{8+2n}   } \frac{  \Gamma[n+4]^2 }{  \Gamma[2n] }    \frac{G m^2}{  \hbar c }   \frac{a^4 \omega^2_{n1} }{ \hbar c},  
 \ee
thus 
\be\label{absor1ton} 
  \CD_g(1s \rightarrow nd)    \approx        \frac{  2^{2n+2} (n-1)^2 n^2   }{  15  (n+1)^{6+2n}   } \frac{  \Gamma[n+4]^2 }{  \Gamma[2n] }    \frac{G  \pi^2}{   c^2 } .    
 \ee
The mean value of the hydrogen radius on the level $nd$ is
\be
\bar{r} = \frac{3}{2}(n^2 -2) a .
\ee
This radius should be smaller than the  mean distance $d$ between hydrogen atoms in the interstellar  gas and the maximal energy level available for  the graviton absorption is defined by the condition  $\bar{r} \sim d$:
\be
n_{max} \sim \sqrt{\frac{2 d}{3 a}  }. 
\ee
In a typical spiral galaxy $d \approx 1cm$, and we will have $n_{max}  \approx 10^8$.  Our approximation for the transition rates is valid  when $\lambda \gg \bar{r}$ and that condition gives $n_{max}  \approx 10^2$. Finally we have
\be 
  \sum^{n_{max}}_{n=3}  \CD_g(1s \rightarrow nd)    \approx      \frac{G  \pi^2}{   c^2 }    \sum^{n_{max}}_{n=3}  \frac{  2^{2n+2} (n-1)^2 n^2   }{  15  (n+1)^{6+2n}   } \frac{  \Gamma[n+4]^2 }{  \Gamma[2n] }       
 \ee
and 
\beqa\label{absorategravtot}
(Total~ Absorption~Rate)_{g}  =   \frac{G \pi^2}{c^2} ~  ~N_{1s} \sum^{n_{max}}_{n=3} ~U(\omega_{n1}) ~ \frac{  2^{2n+2} (n-1)^2 n^2   }{  15  (n+1)^{6+2n}   } \frac{  \Gamma[n+4]^2 }{  \Gamma[2n] } .    
\eeqa
The unknown quantity in these formulas (\ref{absorategrav}), (\ref{absorategrav2}) and (\ref{absorategravtot}) is the spectral energy density of gravitons $U_g(\omega_{n1})d \omega$ at the  angular frequencies  $1.84 \times 10^{16} sec^{-1}  \leq \omega_{n1} \leq  2.1 \times 10^{16} sec^{-1}$ (corresponding to   the wave lengths $91.2~ nm \leq \lambda_{n1} \leq 102.5~ nm$ in Fig.\ref{fig4}).     
We will analyse possible sources of the gravitational radiation  and will estimate the corresponding luminosity of gravitons in the next two sections. It appears that  the graviton luminosity of a typical star  is in the $eV-keV$ range  \cite{Weinberg:1964kqu,Weinberg:1964ew,Weinberg:1965nx,Weinberg:1965rz, Gould}. This energy interval very well overlaps  with this absorption spectrum of the hydrogen atoms. 

However, we would like to stress that  currently, the observational knowledge of the background energy density of gravitons in different regions of the Universe is limited, and here we will consider its value as an unknown quantity that should be determined  by independent astrophysical observations.

\section{\it  Graviton luminosity of stars}

In a series of publications  \cite{Weinberg:1964kqu, Weinberg:1964ew, Weinberg:1965nx, Weinberg:1965rz} Weinberg and Gould \cite{Gould} estimated the power of the thermal gravitational radiation generated by the Sun (or a typical star) through the scattering of electrons and protons $e^- \leftrightarrow e^-$ and $e^- \leftrightarrow p^+$  in a completely ionised hydrogen plasma at the core of the Sun. The thermal graviton frequencies form a continuum spectrum and the produced graviton energy per solid angle $d \Omega$ and per unit  frequency  at  frequency  $\omega$  and direction ${\bf  k}/k$  is \cite{Weinberg:1965nx}:
 \be
 \frac{d E}{d \Omega d \omega}  = \frac{G   (\hbar \omega)^2  }{2 \pi^2  c^5} \sum_{i,j} \frac{\eta_i \eta_j}{( P_i \cdot   k) ( P_j \cdot  k )  } [ (P_i \cdot  P_j)^2 - \frac{1}{2}  (m_i c^2)^2 (m_j c^2)^2], 
  \ee
where $\eta_{in} = -1$ for particles in the initial state, $\eta_{out} = 1$  for particles in the final state and particles 4-momenta are $P_{i}$ and  $ P_j$.
For the nonrelativistic two-body elastic scattering this reduces to 
\be\label{percoll}
 \frac{d E}{ d \omega}  = \frac{8 G \mu^2  }{5 \pi  c^5} v^4 \sin^2 \theta,  
 \ee
where $\mu = m_1m_2/(m_1 +m_2)$  is the reduced mass, $v$ is the relative velocity, and $\theta$  is the scattering angle in the center-of-mass reference frame. For a plasma of completely ionised hydrogen with electron-electron and electron-proton collisions ($e^- \leftrightarrow e^-$ and $e^- \leftrightarrow p^+$) the radiation power per unit volume and per unit frequency interval will be \cite{Weinberg:1965nx}
\be\label{totpercoll}
 \frac{d W_g}{  d \omega}  = \sum_{i,j}\frac{8 G \mu^2_{ij}  }{5 \pi  c^5} \langle v^4_{ij} \int n_i n_j  v_{ij} \frac{d  \sigma_{ij}}{d\Omega}\sin^2 \theta d\Omega \rangle,  
\ee
where $n_i$ is the density of gas particles of type $i$ and $\sigma_{ij}$ is the Rutherford scattering  cross-section. The bracket   $\langle ..... \rangle$  denotes the average over all collisions.  For a typical star the emitted  power  per unit volume and per unit frequency interval is\footnote{The integral over $\theta$ must be cut off at a minimum angle $\theta_0 \ll 1$   determined by Debye screening of the Coulomb force at large impact parameter \cite{Weinberg:1965nx}. Typically $\log(1/\theta_0 )$  is of order 10.}
\be\label{totpercolltyp}
 \frac{d W_g}{ d \omega }  =  \frac{64 G  \hbar^2 }{5  c^2}   \sum_{i,j} n_i n_j   \Big( \frac{2 k T}{\pi  \mu_{ij}  c^2}  \Big)^{1/2} \Big(\frac{e^2_i }{\hbar c}\Big)   \Big(\frac{e^2_j }{\hbar c} \Big) \log(1/\theta_0 ),~   ~~~~
 \ee
where $<v_{ij}> =2 \Big( \frac{2 k T}{\pi  \mu_{ij}}  \Big)^{1/2} $ is the velocity of particles in plasma.  Thus for a plasma of completely ionised hydrogen with electron-electron and electron-proton collisions  (\ref{totpercolltyp})  gives \cite{Weinberg:1965nx}
\be\label{totpercolltypfin}
 \frac{d W_g}{ d \omega }  =  \frac{64 G  \hbar^2 }{5  c^2}   n^2_e   \Big( \frac{2 k T}{\pi  m_{e}  c^2}  \Big)^{1/2}  \Big(\frac{e^2 }{\hbar c} \Big)^2 \log(1/\theta_0 )^{1+\sqrt{2}}  
 \ee
in the units $ \frac{gram }{sec^2 ~ cm}$. The graviton radiation produced by the collisions occurring in a gas (\ref{totpercoll}), (\ref{totpercolltyp}) and (\ref{totpercolltypfin}) is a sum of radiated energies per collision (\ref{percoll}) provided that there is enough time between collisions so that collisions don't interfere, that is,
\be
\omega_c   <  \omega_{g}  < \omega_T,
\ee
where $\omega_c$ is the collision frequency of electrons \cite{Weinberg:1965nx}:
\be
\omega_c  =  \Big(\frac{e^2}{\hbar c}\Big)^2  \Big(\frac{m_e c^2}{k T}\Big)^{3/2}  \Big(\frac{\hbar^2  n_e}{ m^2_e c}\Big),
\ee
and $\omega_T=kT/\hbar$ is the thermal frequency. The spectral energy density is approximately    \be\label{gravspecdens}
U_g  \approx  \frac{d W_g}{\omega_c  d \omega }  = \frac{64 G  m^2  }{5  c}   \Big( \frac{ k T}{  m_{e}  c^2}  \Big)^{2}   n_e    \Big( \frac{\pi}{2}  \Big)^{1/2}     \log(1/\theta_0 )^{1+\sqrt{2}}. 
\ee
Applying these formulas to the hydrogen plasma in the solar core of volume $V_{\odot}  \approx 2 \times 10^{31}~ cm^3$, plasma temperature $T_{\odot} \approx 10^7  K$, electron density $n_e  \approx 3 \times 10^{25} cm^{-3}$  and $\log(1/\theta_0 )  \approx 10$, one can obtain the radiation frequency interval $\triangle \omega_g$ Fig.\ref{fig4}:
\beqa\label{lambdagrange}
&\omega_c =10^{15}~ \sec^{-1} <  \omega_{g} <  \omega_T = 10^{18} ~ \sec^{-1}, \nn\\
&1.88~ nm <  \lambda_{g} <  1884~nm.
\eeqa
The energies  of gravitons $\varepsilon_{g}=\hbar \omega_{g}$ are in the following energy interval:
\be\label{kevrange}
0.66 eV <  \varepsilon_{g} < 658.26 eV .
\ee
The total power produced by the radiation of gravitons  is a product of (\ref{totpercolltypfin}) and $V_{\odot} \triangle \omega_g$. Thus the graviton  luminosity of the Sun is  \cite{Weinberg:1972kfs,Gould}
\beqa\label{sumlumin1}
L_g  \approx   8 \times 10^{14} erg/sec,  
\eeqa
or about $10^{24}$ gravitons per second with energy in the $keV$ range  (\ref{kevrange}).  A recent analysis shows that  \cite{Garcia-Cely:2024ujr}
\beqa\label{sumlumin}
L_g  \approx   1.3 \times 10^{15} erg/sec.
\eeqa
 This comprehensive  analysis of the gravitational emission of the Sun demonstrated that the  partial  contributions consist of:  25\% for photoproduction, 33\%for $ee$ bremsstrahlung, 24\% for $ep$ bremsstrahlung, 17\% for $eHe$ bremsstrahlung, with the rest attributed to the hydrodynamical contribution.

This luminosity provides the energy density of gravitons  on the surface of the Earth:
\be
w_g = \frac{L_g}{4 \pi z^2 c} \approx  9.45 \times 10^{-24} \frac{erg}{cm^3},
\ee
where $z$ is the distance from the Sun. The spectral density    will be approximately    
\be
U = \frac{w_g}{\triangle \omega_g} \approx 5.14 \times 10^{-40} \frac{g}{cm ~sec},
\ee
and the  absorption rate (\ref{absorategrav}), (\ref{absorategravnum3}) would  be 
\beqa
 (Rate~1s \rightarrow 3d)_{g}  =   \Big(\frac{mass~of~H}{10^6 gram}\Big) \times 3.6 \times 10^{-37}  sec^{-1}. 
\eeqa
It is a slow absorption rate, and the hydrogen detector that has the mass of the Earth  would provide the counting rate of $2.2  \times 10^{-15}$ per second, that is,  dozens of gravitons per billions of years \cite{Weinberg:1965nx,Gould,Dyson:2013hbl}.   

There are sources of thermal gravitons that are stronger than the Sun, namely hot white dwarfs at the beginning of their lives and hot neutron stars. The graviton luminosities of a typical white dwarf and a typical neutron star are respectively $10^4$ and $10^{10}$  times solar \cite{Gould}. Their luminosities are roughly proportional to their central densities. The lifetimes during which stars remain hot are shorter than the lifetime of the Sun, being of the order of tens of millions of years for a white dwarf and tens of thousands of years for a neutron star \cite{Gould,Dyson:2013hbl}.   

There are estimates of the magnitude and shape of the gravitational wave background that is produced by the Standard Model physics during the thermal history of the Universe from the primordial thermal plasma, the Cosmic Gravitational Microwave Background (CGMB) \cite{Ghiglieri_2015, Ringwald_2021}.   

The source of gravitational waves  in $100~kHz$ frequency range, corresponding to the transition between states adjacent  to $ n_{max} \approx   10^{4}$, induced by the annihilation of QCD axions in the cloud they may form around stellar mass black holes were discussed in \cite{PhysRevLett.110.071105}. The relic gravitons can also be created from zero-point quantum fluctuations in the course of the cosmological expansion with the estimated frequencies in the range $10^{-18} - 10^{-16} Hz$ \cite{Grishchuk_1989} that would  corresponds  to the transition between adjacent quantum states with $n_{max} \approx 10^{10}$.  

{\it However,  currently the observational knowledge of the background energy density of gravitational radiation in different regions of the Universe is limited, and here we will consider its value as an unknown quantity that should be determined  by independent astrophysical observations.  Our aim is to suggest a possible method for measuring the intensity of the gravitational radiation in different regions of space. }

It seems natural to turn attention to the clouds of  interstellar  hydrogen and to the giant halos of pristine, primordial hydrogen gas that surrounded galaxies in the early universe  at z = 5.5 - 13.4. The presents of copious amount of pristine HI gas was discovered  by measurement of  the  strong Ly$\alpha$ emission and  absorption line in a large spectroscopic sample of star-forming galaxies at $z = 5.5 - 13.4$ observed with JWST\cite{2023A&A...677A..88B,  2025A&A...693A..60H,  Kashino_2023,heintz2023,  2024ApJ...962...24S}.


A typical galaxy contains a vast amount of hydrogen, which is the most abundant element.   Hydrogen in galaxies exists as a neutral atomic hydrogen $HI$,  cold  molecular hydrogen $H_2$  in dense clouds and as a highly ionised  $HII$ hydrogen near hot stars.  The neutral atomic hydrogen $HI$  presents a significant amount of the cold diffused medium that is filling much of the galaxies.   The interstellar medium  (ISM) makes up roughly 5\% galaxy's total mass, and about 70\% of that gas is hydrogen, and most of the hydrogen atoms are in the ground  state \cite{Dyson2020}. Their mass can be\footnote{A typical galaxy has hundreds of billions of stars, and each star is primarily hydrogen.   Individual stars are composed of roughly 70-75\% hydrogen by mass during their lifetime. A significant portion of hydrogen  exists in the form of interstellar gas clouds, which serve as the fuel for new star formation.  Overall, the vast majority of all ordinary matter in a typical galaxy is hydrogen. } 
\be
M^{H}_{gal} \approx 10^{44}  gram, 
\ee
which amounts to
\be
N_{1s} = M^{H}_{gal} ~ N_A  \approx 10^{68} 
\ee
hydrogen atoms. The graviton  luminosity of a typical galaxy $L_{gal}$ can be obtained by taking the luminosity $L_g$ (\ref{sumlumin}) times the average number of stars in galaxy. In that case the graviton luminosity $L_{gal}$ of a typical galaxy  of hundred billions stars $10^{11}$   can be  
\beqa
L_{gal} \approx   L_g \times 10^{11} = 1.3 \times 10^{26} erg/sec.
\eeqa
While for a typical  white dwarf $L_{WD}= 10^4 L_g$ and  for a typical neutron star $L_{NS}= 10^{10} L_g$ \cite{Gould},  therefore the graviton luminosity can be between $10^{26} - 10^{36} erg/sec$.

The conclusion that can be drown from this consideration is  that  the absorption rate of gravitons by galactic hydrogen atoms can be quite large and  our aim is to investigate  methods that will allow  to measure the intensity of graviatational radiation.   It could be advantageous to compare the photon luminosities of  clouds of hydrogen atoms that are coming from different regions of spiral or elliptic galaxies where the photon radiation background is minimal.  In the  next section we will consider the photon background luminosity induced by stars and compare it with their  graviton luminosity. The ratio $R_{\gamma}$ of the photon luminosities will be defined that is maximally sensitive to the absorption of gravitational radiation. 

\section{\it   Photon luminosity of stars}

\begin{figure}
 \centering
\includegraphics[angle=0,width=9cm]{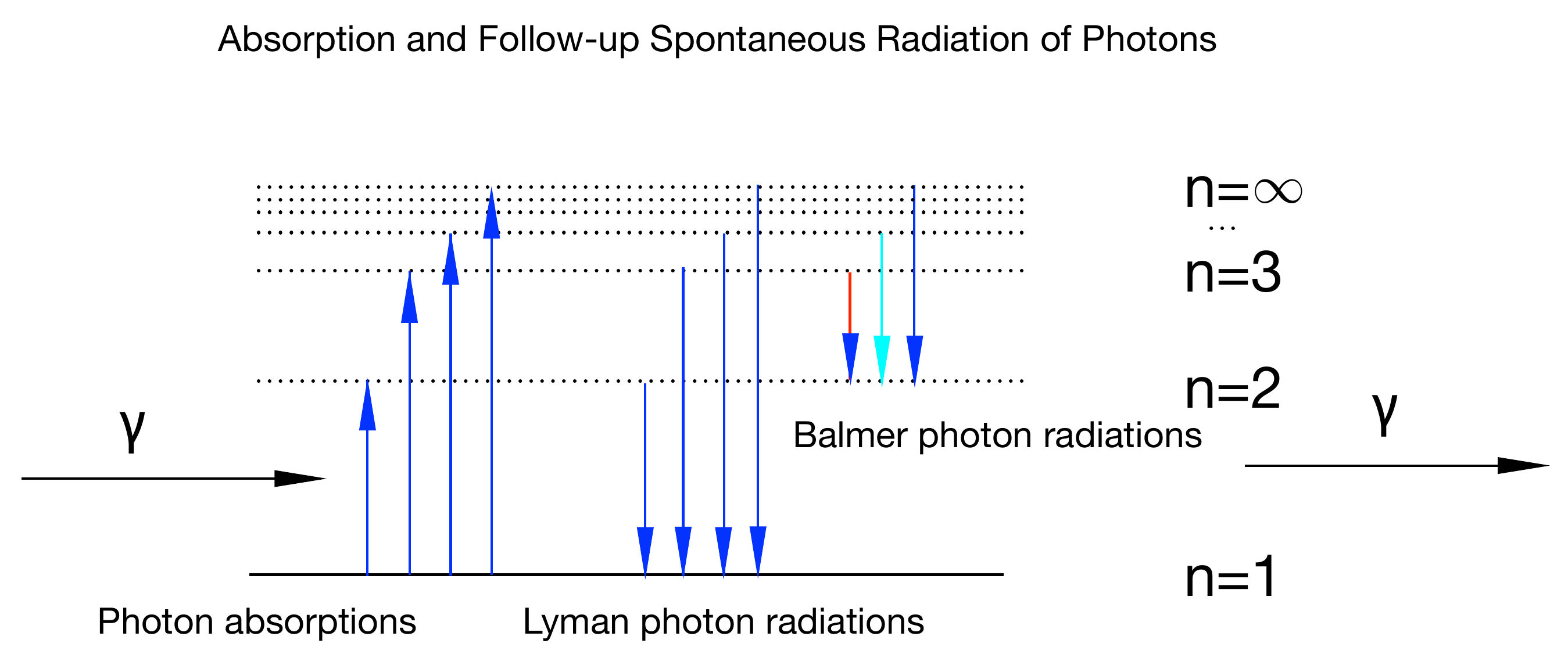} ~~
\centering
\caption{The figure demonstrates the absorption of photons ($\gamma$) and the follow-up  spontaneous radiation.   The spectral density of the background photons is defined by $U(\omega,T)$ (\ref{equa5}).
The Lyman photons radiation wave lengths are in the UV region $91.2~nm \leq \lambda_{Ly} \leq 121.6 ~nm$, while the wave lengths of the Balmer photons are in the visible part of the spectrum $364.6~nm \leq \lambda_H \leq 656.2 ~nm$  \cite{Kramida}.}
\label{fig2} 
\end{figure}

The surface of the Sun  emits electromagnetic radiation across a broad spectrum, and its emission is peaked in the visible and infrared region of the spectrum \cite{Dyson2020, 2004SoEn...76..423G}.  The spectral energy density (energy per unit volume and per unit frequency) of the Sun can be approximated  by the Planck formula of the black body radiation with the effective temperature at about $ T_{\odot} = 5800 K$:
\be\label{equa5}
 U_{\gamma}(\omega,T) = \frac{\hbar \omega^3}{\pi^2  c^3}\frac{1}{e^{\frac{\hbar \omega}{k T}    }-1}.
 \ee
The photon luminosity of the Sun yields 
 \be
 L_{\gamma} = 4 \pi R^2_{_{\odot}} \int \frac{\hbar \omega^3}{\pi^2  c^2}\frac{d \omega}{e^{\frac{\hbar \omega}{k T}    }-1} ,
 \ee
where $R_{_{\odot}}$ is the Sun radius. This luminosity is approximately  
\be
 L_{\gamma}  \approx 3.83 \times 10^{33} \frac{erg}{sec}.
\ee
The ratio of graviton luminosity (\ref{sumlumin}) to the photon luminosity in the vicinity of the Sun surface is
\be
\frac{L_{g}}{L_{\gamma}} \approx  10^{-18}
\ee
and is an {\it  extremely  small quantity considered in the vicinity of a  star}. This ratio can be improved by the consideration of {\it specific energy bands and regions of spiral galaxies, where the photon luminosity is minimal due to the absorption of photons at the boundaries of the hydrogen and dust clouds}. In particular,  it is advantageous that the graviton absorption rate to the $n \geq 3$ states is in the Lyman ultraviolet (UV) region $91.1~nm \leq \lambda_g \leq 102.5 ~nm$, where the photon radiation spectrum in UV region (\ref{equa5})  is less intensive  \cite{2004SoEn...76..423G} (see Fig. \ref{fig2}).  Stars in a galaxy may have masses within the range of about $0.1$ to about $100$ times the mass of the Sun, and the luminosities $ L_{\gamma} $ corresponding to these masses may range from about $10^{-3}$ to $10^6$ times the luminosity of the Sun \cite{Dyson2020}, thus the observation of the regions with stars that have low masses would be preferable  
 $$
\frac{l_{g}}{L_{\gamma}}   \approx   10^{-15} .
$$
In the next section we suggest measuring the ratio of photon luminosities that is  defined in a way  that maximally exposes the fundamental difference in the nature of photons and gravitons - their helicity, which is $h=1$ for the  photons and $h=2$  for the gravitons \cite{Savvidy:2025rqt}.

\section{\it  The ratio  of photon  luminosities }

\begin{figure}
 \centering
\includegraphics[angle=0,width=10cm]{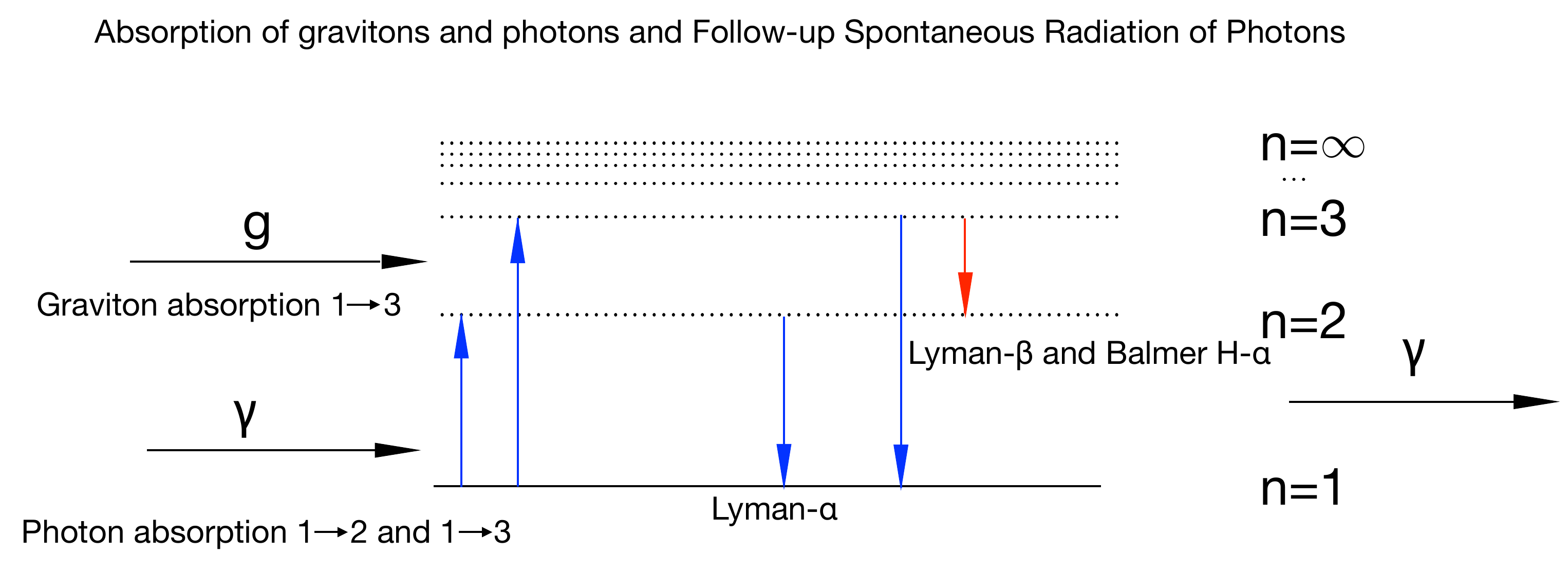} ~~
\centering
\caption{The absorption of gravitons to the $n=2$ state vanishes  (\ref{zerotrans}), while the photons absorption  to the $n=2$ state  is not. Therefore a spontaneous radiation of photons from  $n=2$ state has a contribution only from the photon absorption, while the spontaneous radiation   from $n=3$ has contributions from the absorptions of gravitons and photons.  This fundamental fact can be used to identify and measure the excess  of photons from the $n=3$ state that should be attributed to the gravitons.}
\label{fig3} 
\end{figure}
The transition rate of gravitons to the $n=2$ state  identically vanishes  because of its helicity  (\ref{zerotrans}), while the absorption rate of the photons to the same state (Lyman-$\alpha$) is not vanishing.  Therefore the spontaneous radiation of photons from  $n=2$ state is induced only by the absorption of photons (see Fig. \ref{fig3}).  In contrast,  the spontaneous radiation of photons from $n=3$ has the contributions from the absorption of both quanta: the photons  and the gravitons. This difference in the origin of photons that are spontaneously radiated from  $n=2$ and $n=3$ states can be used to measure the excess  of photons from the $n=3$ state compared with the $n=2$ state.  

We suggest a method for measuring the excess of the photons by measuring the corresponding luminosity ratio. This ratio can be computed in the following way.  The number of  hydrogen atoms that are exited to the second $n=2$ and to the third  $n=3$ energy  levels by absorption of photons during the time interval $\triangle t$   are  (the solutions of the rate equation is discussed in Appendix C) 
\beqa
&&N_{2\gamma} = \CD_{\gamma}(1 \rightarrow 2) U_{\gamma}(\omega_{21}) \triangle t, \\
&&N_{3\gamma} = \CD_{\gamma}(1 \rightarrow 3) U_{\gamma}(\omega_{31}) \triangle t +\CD_{\gamma}(2 \rightarrow 3) U_{\gamma}(\omega_{32}) \triangle t, \nn
\eeqa
where the photon absorption rate is \cite{Dirac:1927dy,Fermi:1932xva, Fermi1954} (Appendix B contains  tables of the photon absorption rates)
\be\label{photonabsorf1}
 \CD_{\gamma} (m \rightarrow n) = \frac{4 \pi^2 e^2}{3 \hbar^2} \vert \langle n \vert  x_i \vert m  \rangle \vert^2.
\ee
The luminosity emitted from these energy levels can be computed in the following form:
\beqa
L^{\gamma}_{21} = \hbar \omega_{21}  N_{2\gamma} ~ \CS_{\gamma}(2 \rightarrow 1),\nn\\
L^{\gamma}_{31} = \hbar \omega_{31}  N_{3\gamma}  ~\CS_{\gamma}(3 \rightarrow 1),\nn\\
L^{\gamma}_{32} = \hbar \omega_{32}  N_{3\gamma}  ~\CS_{\gamma}(3 \rightarrow 2), 
\eeqa
where the photon emission rate is \cite{Dirac:1927dy,Fermi:1932xva,Fermi1954}
\be\label{fermiabs}
 \CS_{\gamma} (m \rightarrow n) = \frac{4 e^2 \omega^3_{mn}}{3 ~\hbar c^3} \vert \langle n \vert  x_i \vert m  \rangle \vert^2 .
\ee
The ratio that could  play an important role in the measuring the graviton intensity  is  the ratio of the photon luminosities that are emitted from the second and the third energy levels\footnote{The luminosities $L^{\gamma}_{21} \equiv L(Ly_{\alpha})$  and   $L^{\gamma}_{21} \equiv L(Ly_{\beta})$  are the Lyman lines luminosities, while   $L^{\gamma}_{32} \equiv L(H_{\alpha})$  is the Balmer line luminosity.}:
\be\label{basrat}
\CR_{\gamma} = \frac{L^{\gamma}_{31} + L^{\gamma}_{32}}{L^{\gamma}_{21} }.
\ee
 This ratio is based on the selection rule that prevents the absorption of particles of helicity $h=2$ to the second energy level  of the hydrogen atoms (\ref{zerotrans}):
$$
 \CD_g(1s \rightarrow 2)  =   0.
$$
The number of  the hydrogen atoms that are exited during the time interval $\triangle t$     to the  third energy level of the hydrogen atom by the absorption of gravitons is (\ref{fundabsorbrate})
\beqa
 N_{2g} =0,~~~~~~~~ N_{3g} = \CD_{g}(1 \rightarrow 3) ~U_{g}(\omega_{31}) \triangle t,
\eeqa
and we have (\ref{fermiabs})
\beqa
L^{g}_{21} = 0,~~~~~~
L^{g}_{31} = \hbar \omega_{31}  N_{3g} ~ \CS_{g}(3 \rightarrow 1),~~~~~~
L^{g}_{32} = \hbar \omega_{32}  N_{3g} ~ \CS_{g}(3 \rightarrow 2).
\eeqa
One can obtain the explicit  expression for the dimensionless ratio of the luminosities  (\ref{basrat}) by substituting the expressions of the absorption and emission rates 
\be
\CR_{\gamma} = C_{31}~\frac{e^{\frac{\hbar \omega_{21}}{k T}}  - 1}{e^{\frac{\hbar \omega_{31}}{k T}}  - 1} +C_{32}~  \frac{e^{\frac{\hbar \omega_{21}}{k T}}  - 1}{e^{\frac{\hbar \omega_{32}}{k T}}  - 1}, 
\ee
where (see Appendix B)
\be
C_{31} = \frac{3^{16} ~ 11~ 821}{2^{21} ~ 5^9}, ~~~~~  C_{32}=\frac{3^{16}~ 11 ~23~ 41~ 821}{2^{8} ~ 5^{19}} .
\ee
The dimensionless ratio of the luminosities  (\ref{basrat}) can be expressed as the ratio of emission rates, which  is a pure rational number (see Appendix B)
\be\label{threelevelcase}
(Emission ~Ratio) =\frac{ \omega_{31} \CS_{\gamma}(3 \rightarrow 1)  + \omega_{32} \CS_{\gamma}(3 \rightarrow 2) }{\omega_{21} \CS_{\gamma}(2 \rightarrow 1) }  = \frac{821 \times 11 \times 3^{9}}{2^{8} 5^{9}}  \approx 0.355514 
\ee
and  of the absorption rates 
\be\label{absbasrat}
(Absorp.~ Ratio) = \frac{N_{3\gamma}}{ N_{2\gamma} }= \frac{\CD_{\gamma}(1 \rightarrow 3) U_{\gamma}(\omega_{31}) +\CD_{\gamma}(2 \rightarrow 3) U_{\gamma}(\omega_{32})}{\CD_{\gamma}(1 \rightarrow 2) U_{\gamma}(\omega_{21})} \approx 3.62614 \times 10^{6},
\ee
that essentially depends on the intensity of the background radiation (\ref{equa5}) (see Appendix B (\ref{tempabsodep}))
Thus, for  the Sun the numerical value of  the  $\CR_{\gamma}$ would be 
\be\label{mainnumber}
 \CR_{\gamma}=\frac{L^{\gamma}_{31} + L^{\gamma}_{32}}{L^{\gamma}_{21} } \approx 1.28915 \times   10^{6}.
\ee
This number is much bigger than the observational  value. The corresponding luminosities of the Sun are \cite{Delbouille, 1996ApOpt..35.2747A, Asplund_2009}:
\be
L^{\gamma}_{21}  \approx 2 \times 10^{28}  erg/sec,~~ L^{\gamma}_{31}  \approx 5 \times 10^{26}  erg/sec,~~L^{\gamma}_{32}  \approx 1.9 \times 10^{30}  erg/sec, 
\ee
and we get 
$$
R_{\gamma} \approx 95.
$$ 
The difference appears because the luminosities $L^{\gamma}_{21} $ and $L^{\gamma}_{31} $  of the $Ly_{\alpha}$ and $Ly_{\beta}$ photons are generated in the Sun chromosphere and transient layer at temperature $ 10^4 - 10^6 K$, while the luminosity $L^{\gamma}_{32} $ of Balmer photons  $H_{\alpha}$ is generated in the Sun photosphere at temperature  $5800 K$ \cite{refId0}. With this correction to the spectral densities $U(\omega_{21})$ and $U(\omega_{31})$ we obtain  $\CR_{\gamma}\approx 81$. 

Recent JWST spectroscopic observations  allow  to study the most distant galaxies at $z \approx 11-13$ and  the properties of the neutral, pristine HI gas that surrounding  the first galaxies.  The presents of copious amount of pristine HI gas was discovered by measurement of  the  strong Ly$\alpha$ emission and  absorption line in a large spectroscopic sample of star-forming galaxies at $z = 5.5 - 13.4$ observed with JWST\cite{2023A&A...677A..88B,  2025A&A...693A..60H,  Kashino_2023,heintz2023,  2024ApJ...962...24S,ning2026}. The information concerning much needed luminosities of the $Ly_{\alpha}$, $Ly_{\beta}$ and $H_{\alpha}$ spectral lines can be found in recent publications on the JWST spectroscopic observations.  Only the ration $H_{\alpha}/Ly_{\alpha}$ can be computed from available data, because $Ly_{\beta}$ is blanked \cite{heintz2023},  and it is in the range $0.27 - 0.39$ \cite{ ning2026}.   A dedicated analysis of the  observational data from the  JWST  will be  important in measuring these luminosities.

Thus, in order to measure the deviation of the photon radiation  from  interstellar  and intergalactic regions induced by gravitons  it is important to know the theoretical ratio $\CR_{\gamma}$ as precise as possible.  The above consideration demonstrated that the spectral luminosities should be measured from the regions   that have the most uniform distribution of temperature. The evaluation  of the $\CR_{\gamma}$ to  the  form that is most convenient for the calculation  (\ref{mainratio}) is given in the  Appendix B.

\section{\it Acknowledgement }   
  
One of us, G.S., would like to thank Prof. Pengming Zhang for the invitation and kind hospitality at the Sun Yat-Sen University where part of this work was conducted.

\begin{appendices}

\section{\it  Wave functions and graviton emission rates}

\beqa\label{append}
&&R_{10}=\frac{2 e^{-\frac{r}{a}}}{\sqrt{a^3}}, ~  R_{20}=\frac{e^{-\frac{r}{2 a}} }{\sqrt{2 a^3}} \left(1-\frac{r}{2 a}\right),~~ 
  R_{21}=\frac{ e^{-\frac{r}{2 a}}}{ 2 \sqrt{6 a^3}  } \left(\frac{r}{a}\right),\\
&&R_{30}=\frac{2 e^{-\frac{r}{3 a}} }{3 \sqrt{3 a^3}} \left(1 - \frac{2 r}{3 a} + \frac{2}{27} \left(\frac{r}{a}\right)^2\right),~R_{31}=\frac{8   e^{-\frac{r}{3 a}}}{ 27 \sqrt{6 a^3} } \left(\frac{r}{a}\right)\left(1-\frac{r}{6 a}\right), \nn\\ 
&&~R_{32}=\frac{4 e^{-\frac{r}{3 a}}}{81 \sqrt{30 a^3}}  \left(\frac{r}{a}\right)^2, ~~
R_{n2} \approx \frac{ (-1)^{n-3} 2^n }{n^{n+1} \sqrt{\Gamma[2n] a^3}} e^{-\frac{r}{n a}} \left(\frac{r}{a}\right)^{n-1}  ~~~~r \gg a ,\nn   \\
&&  R_{n2} \approx \frac{ 1 }{ 15 n^4 }  \sqrt{  \frac{(n+2)!}{ (n-3)!  a^3}  }   \left(\frac{r}{a}\right)^{2}  ~~~~r \approx a  .\nn
\eeqa
The first nonzero radiation of gravitons  appears from the $\vert 3d \rangle$ to the $\vert 1s \rangle$ state and the corresponding matrix elements  are defined as
\beqa\label{matrixelem}
 &&\langle 3d_{l_z} \vert D_{ij} \vert 1s \rangle = m \int   R_{32} Y_{2,l_z}  ~  (3 x_{i} x_{i} - \delta_{ij} r^2)   R_{10} Y_{0,0} ~r^2 dr \sin \theta d\theta d \phi= \CD_{l_z}(ij),~~~\nn\\
&& l_z =0,\pm1,\pm2~~~~~i,j=1,2,3, 
 \eeqa
 where the radial wave functions $R_{nl}$ are given above. By calculating these matrices one can get:
 \beqa\label{matquanra}
&&\CD_2(ij) =  \CD_{-2}(ij)^* =  \frac{243 }{256}\begin{pmatrix}
1 & i & 0 \\
i& -1 & 0 \\
0 & 0 &0
\end{pmatrix}m a^2,~\nn\\
&&\CD_1(ij) =  - \CD_{-1}(ij)^*=  \frac{243}{256}\begin{pmatrix}
0 & 0 & -1 \\
0& 0 & -i \\
-1 & -i &0
\end{pmatrix}m a^2,~\\
&&\CD_0(ij) =  \frac{81 \sqrt{3}}{128 \sqrt{2}}\begin{pmatrix}
-1 & 0 & 0 \\
0& -1 & 0 \\
0 & 0 &2
\end{pmatrix}m a^2,~\nn
\eeqa
where   $a=  \hbar^2/m e^2$ is the Bohr radius and $m$ is the mass of the electron.  All other matrix elements $\langle 3p \vert D_{ij} \vert 1s \rangle$ and $\langle 3s \vert D_{ij} \vert 1s \rangle$ vanish. 
For the matrix elements  $\langle 4d \vert D_{ij} \vert 1s \rangle$  we obtain:
 \beqa\label{matquanra2}
&&\CD_2(ij) =  \CD_{-2}(ij)^* =  \frac{98304 \sqrt{6}}{390625}\begin{pmatrix}
1 & i & 0 \\
i& -1 & 0 \\
0 & 0 &0
\end{pmatrix}m a^2,~\nn\\
&&\CD_1(ij) =  - \CD_{-1}(ij)^*=  \frac{98304 \sqrt{6}}{390625}\begin{pmatrix}
0 & 0 & -1 \\
0& 0 & -i \\
-1 & -i &0
\end{pmatrix}m a^2,~\\
&&\CD_0(ij) =  \frac{198608 }{39625 }\begin{pmatrix}
-1 & 0 & 0 \\
0& -1 & 0 \\
0 & 0 &2
\end{pmatrix}m a^2.~\nn
\eeqa
All other matrix elements $\langle 4s \vert D_{ij} \vert 1s \rangle$ and $\langle 4f \vert D_{ij} \vert 1s \rangle$ vanish.

\section{\it Ratio of photon luminosities }


The spontaneous radiation of photons  from  all the $n \geq 3$ states has a contribution from absorption of background photons and gravitons. The spontaneous radiation of gravitons from the $n \geq 3$ states (\ref{grtrans1}), (\ref{grtrans2}) and (\ref{grtransn})  is a much slower process and can be ignored compared with the photon radiation, which is given by the expression (\ref{fermiabs}).   The total spontaneous transition rates for the photons are\footnote{The total transition rate is a sum of partial transition rates over all final and initial states.} 
 \beqa\label{balmprob}
&\CS_{\gamma}(2p \rightarrow 1s) = \frac{2^9}{3^7} \Big(\frac{e^2}{\hbar c}\Big)^3   \Big(\frac{m e^4}{2  \hbar^3}\Big) ,~~~~  \CS_{\gamma}(3p \rightarrow 1s) = \frac{1}{2^4} \Big(\frac{e^2}{\hbar c}\Big)^3   \Big(\frac{m e^4}{2  \hbar^3}\Big) , \nn\\
&\CS_{\gamma}(3d \rightarrow 2p) = \frac{3\  2^{17}}{ 5^{10}} \Big(\frac{e^2}{\hbar c}\Big)^3   \Big(\frac{m e^4}{2  \hbar^3}\Big),~~~~ 
\CS_{\gamma}(3p \rightarrow 2s) = \frac{2^{14}}{5^{9}} \Big(\frac{e^2}{\hbar c}\Big)^3   \Big(\frac{m e^4}{2  \hbar^3}\Big),~\nn\\
&\CS_{\gamma}(3s \rightarrow 2p) = \frac{3~2^9 }{5^9} \Big(\frac{e^2}{\hbar c}\Big)^3   \Big(\frac{m e^4}{2  \hbar^3}\Big), ......
 \eeqa
 \begin{figure}
 \centering
\includegraphics[angle=0,width=9cm]{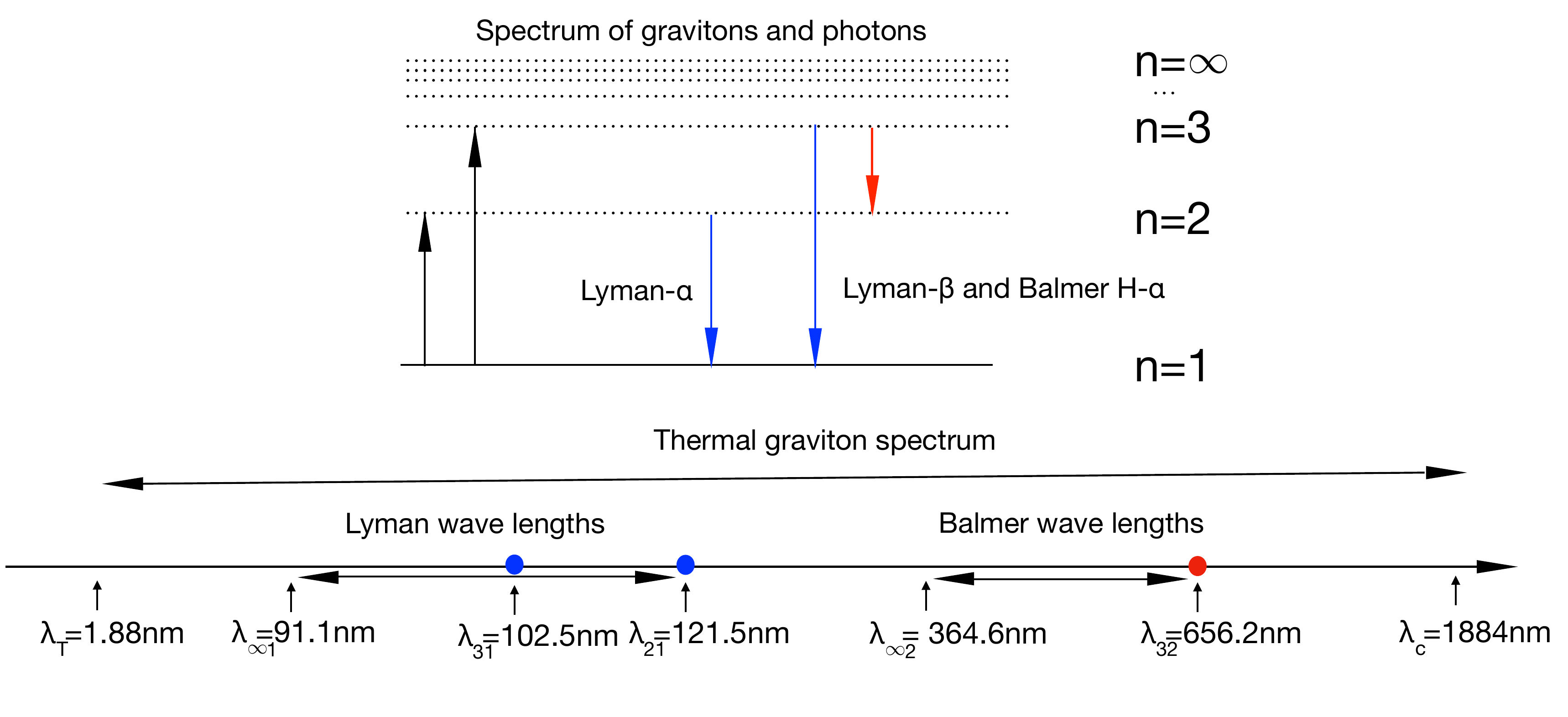} ~~
\centering
\caption{The spontaneous radiation  at $\lambda_{21}=121.5~nm$ is induced exclusively  by the absorption of photons, while the spontaneous radiation  at $\lambda_{31}= 102.5~ nm$ and $\lambda_{32}= 656.2~ nm$ is induced by the absorption of photons and of gravitons.   The measurement of the difference in the intensities at these wave lengths can be used to identify the absorption of gravitons.}
\label{fig4} 
\end{figure}
 and have the following magnitudes:   
 \beqa
 & (2p \rightarrow 1s) =1.882 \times 10^{9} \frac{1}{sec},~~~~
 (3p \rightarrow 1s) =5.024 \times 10^{8} \frac{1}{sec},\nn\\
 &(3d \rightarrow  2p) =3.237 \times 10^{8} \frac{1}{sec},~~~~ 
 (3p \rightarrow 2s) =6.743 \times 10^{7} \frac{1}{sec},~  \nn\\
&(3s \rightarrow 2p) =6.32 \times 10^{6} \frac{1}{sec},.....~~~~
\eeqa  
The emission ratio (\ref{threelevelcase}) has the following form 
\beqa
&&~~~~~~~~~~~~~~~~~~~~~~~~~~~(Emission ~Ratio) =\\
&&=\frac{ \omega_{31}\CS_{\gamma}(3p \rightarrow 1s)   + \omega_{32} (\CS_{\gamma}(3d \rightarrow 2p) +\CS_{\gamma}(3p \rightarrow 2s) +\CS_{\gamma}(3s \rightarrow 2p)) }{\omega_{21} \CS_{\gamma}(2p \rightarrow 1s) } \nn
\eeqa
and can be evaluated using the table of the photon spontaneous transition rates (\ref{balmprob})  and the corresponding frequencies  (\ref{frequ})
\be\
\omega_{21}=  \frac{3}{4}  \frac{m e^4}{2 \hbar^3} ,~~~~\omega_{31}=  \frac{8}{9}  \frac{m e^4}{2 \hbar^3},~~~~\omega_{32}=  \frac{5}{36}  \frac{m e^4}{2 \hbar^3} , 
\ee
thus
\beqa
(Emission ~Ratio) =\frac{ \frac{8}{9}  \frac{1}{2^4}    +  \frac{5}{36}  ( \frac{3\  2^{17}}{ 5^{10}}  + \frac{2^{14}}{5^{9}} +\frac{3~2^9 }{5^9}) }{ \frac{3}{4}\frac{2^9}{3^7}}  = 
\frac{821 \times 11 \times 3^{9}}{2^{8} 5^{9}}. 
\eeqa
The absorption rate of the photons is defined by the expression (\ref{photonabsorf1}) and has the following form:
\beqa\label{photonabsor1}
& \CD_{\gamma}(1s \rightarrow 2p) = \frac{ 2^{17}}{3^{10}} \Big(\frac{e^2}{\hbar c}\Big)^{-1}   \Big(\frac{ \pi^2 \hbar }{ m^2 c}\Big) ,~~~~   \CD_{\gamma}(1s \rightarrow 3p) = \frac{3^6}{2^{11}}  \Big(\frac{e^2}{\hbar c}\Big)^{-1}   \Big(\frac{ \pi^2 \hbar }{ m^2 c}\Big), \nn\\
& \CD_{\gamma}(2p  \rightarrow 3d) = \frac{3^7 2^{25}}{5^{13}}  \Big(\frac{e^2}{\hbar c}\Big)^{-1}   \Big(\frac{ \pi^2 \hbar }{ m^2 c}\Big),~~~~ 
 \CD_{\gamma}(2s \rightarrow 3p ) = \frac{3^6 2^{22}}{5^{12}}  \Big(\frac{e^2}{\hbar c}\Big)^{-1}   \Big(\frac{ \pi^2 \hbar }{ m^2 c}\Big) ,~\nn\\
& \CD_{\gamma}(2p  \rightarrow 3s) = \frac{3^7 2^{17}}{5^{12}}  \Big(\frac{e^2}{\hbar c}\Big)^{-1}   \Big(\frac{ \pi^2 \hbar }{ m^2 c}\Big), ......
 \eeqa
with the following numerical values:
 \beqa
& \CD_{\gamma}(1s \rightarrow 2p)  \approx 1.27 \times 10^{20} \frac{cm}{gram},~~~~~~ \CD_{\gamma}(1s \rightarrow 3p)  \approx 2.04 \times 10^{19} \frac{cm}{gram}, \nn\\
& \CD_{\gamma}(2p \rightarrow 3d)  \approx 3.45 \times 10^{21} \frac{cm}{gram},~~~~~~  \CD_{\gamma}(2s \rightarrow 3p)  \approx 7.18 \times 10^{20} \frac{cm}{gram},  \nn\\
&\CD_{\gamma}(2p \rightarrow 3s)  \approx 6.73 \times 10^{19} \frac{cm}{gram},....
 \eeqa
The absorption ratio  (\ref{absbasrat}) has the following form 
\be
\frac{N_{3\gamma}}{ N_{2\gamma} }= \frac{\CD_{\gamma}(1s \rightarrow 3p) U_{\gamma}(\omega_{31}) +(\CD_{\gamma}(2p  \rightarrow 3d)+   \CD_{\gamma}(2s \rightarrow 3p ) +  \CD_{\gamma}(2p  \rightarrow 3s) ) U_{\gamma}(\omega_{32})}{\CD_{\gamma}(1s \rightarrow 2p)   U_{\gamma}(\omega_{21})}  
\ee
and can be evaluated by using the table of photon absorption rates (\ref{photonabsor1})  and the Planck black body radiation density   (\ref{equa5})  of the astronomical object under consideration. Thus
\be\label{tempabsodep}
\CR_{\gamma}=   C_{31}   \frac{^{\frac{\hbar \omega_{21}}{k T}    }-1}{e^{\frac{\hbar \omega_{31}}{k T}    }-1}  
+
C_{32}   \frac{^{\frac{\hbar \omega_{21}}{k T}    }-1}{e^{\frac{\hbar \omega_{32}}{k T}    }-1} , 
\ee
where 
\be
C_{31} =  \frac{ \frac{3^6}{2^{11}}   (\frac{8}{9})^3  }{    \frac{ 2^{17}}{3^{10}}  (\frac{3}{4})^3   } , ~~~
C_{32} = \frac{   (\frac{3^7 2^{25}}{5^{13}}+ \frac{3^6 2^{22}}{5^{12}}   +  \frac{3^7 2^{17}}{5^{12}}) (\frac{5}{36})^3  }{\frac{ 2^{17}}{3^{10}}  (\frac{3}{4})^3      }.
\ee

 \section{\it Rate equation}  
 
 The rate equation for the three lower states has the following form  
 \beqa\label{ratesystem}
\frac{d N_{2\gamma}}{d t} &= &\CD_{\gamma}(1  \rightarrow 2) U_{\gamma}(\omega_{21}) N_{1s} +\Big( \CS_{\gamma}(3  \rightarrow 2)  +\CD_{\gamma}(3  \rightarrow 2) U_{\gamma}(\omega_{32})   \Big) N_{3\gamma} \nn\\
&-&\Big( \CS_{\gamma}(2  \rightarrow 1)  +\CD_{\gamma}(2  \rightarrow 1) U_{\gamma}(\omega_{21}) +\CD_{\gamma} (2  \rightarrow 3)    U_{\gamma}(\omega_{32})  \Big) N_{2\gamma} \nn\\ \\
\frac{d N_{3\gamma}}{d t} &= &\CD_{\gamma}(1  \rightarrow 3) U_{\gamma}(\omega_{31}) N_{1s} +   \CD_{\gamma} (2  \rightarrow 3)    U_{\gamma}(\omega_{32})  N_{2\gamma} \nn\\
&-&\Big( \CS_{\gamma}(3  \rightarrow 1)  +\CD_{\gamma}(3  \rightarrow 1) U_{\gamma}(\omega_{31})   + \CS_{\gamma}(3  \rightarrow 2)      +\CD_{\gamma}(3  \rightarrow 2) U_{\gamma}(\omega_{32})     \Big) N_{3\gamma},  \nn
\eeqa
where the $N_{2\gamma}$ and $N_{3\gamma}$ are the excited-state populations. The tables of absorption and emission rates are  given in Appendix B.  The energy density of the photon radiation $U_{\gamma}$ depends on a given environment. The coefficients on the right hand side of this system have different orders of magnitude and the main contribution in case of clouds ISM/IGM comes from the term proportional to $N_{1s}$.  Therefore a dominant contribution has the following form  
\beqa
&& N_{2\gamma} \sim \CD_{\gamma}(1 \rightarrow 2) U_{\gamma}(\omega_{21}) N_{1s}~\triangle t, \\
&&  N_{3\gamma} \sim  \CD_{\gamma}(1 \rightarrow 3) U_{\gamma}(\omega_{31}) \triangle t +\CD_{\gamma}(2 \rightarrow 3) U_{\gamma}(\omega_{32}) N_{1s} ~\triangle t, \nn
\eeqa
 and will enter into the ratio $\CR_{\gamma}$ as 
 \beqa
 \frac{  N_{3\gamma}}{  N_{2\gamma} } = \frac{\CD_{\gamma}(1 \rightarrow 3) U_{\gamma}(\omega_{31}, T) +\CD_{\gamma}(2 \rightarrow 3) U_{\gamma}(\omega_{32}, T )}{\CD_{\gamma}(1 \rightarrow 2) U_{\gamma}(\omega_{21}, T )} 
 \eeqa
The definition of the ratio
$$ 
 \CR_{\gamma}=\frac{L^{\gamma}_{31} + L^{\gamma}_{32}}{L^{\gamma}_{21} } 
 $$  includes the contributions of the main transitions rates. Because the gravitons can also be absorbed to the high energy levels of hydrogen atom, the definition of $\CR_{\gamma}$ should  include the absorption and emission rates of photons from  high energy states as well
\beqa
\CR_{\gamma} = \frac{ L^{\gamma}_{Lyman} +L^{\gamma}_{Balmer} +L^{\gamma}_{Paschen} +...}{L^{\gamma}_{21} }. 
\eeqa
In order to measure the deviation of the photon radiation  from  interstellar  and intergalactic regions induced by a possible absorption of helicity two particles,  it is important to know the theoretical ratio $\CR_{\gamma}$ as precise as possible and  that should be done by solving the rate equation (\ref{ratesystem}) in a given environment.

\section{\it Stability of atoms} \label{concl}

The existence of the gravitational waves was predicted in two articles written in 1916 \cite{Einstein1} and in 1918 \cite{Einstein2} by  Einstein.  The  formula  obtained in these articles describes the intensity of gravitational waves that are generated by accelerating bodies.  In the concluding part of the first article \cite{Einstein1} Einstein raised the question of the stability of atoms due to the radiation of gravitational waves. Einstein wrote: 

"Gleichwohl müßten die Atome zufolge der inneratomischen Elektronenbewegung nicht nur elektromagnetische, sondern auch Gravitationsenergie ausstrahlen, wenn auch in winzigem Betrage. Da dies in Wahrheit in der Natur nicht zutreffen dürfte, so scheint es, daß die Quantentheorie nicht nur die Maxwellsche Elektrodynamik, sondern auch die neue Gravitationstheorie wird modifizieren müssen."

This concluding remark says that due to the internal motion of electrons in atoms the atoms would have to emit not only electromagnetic but also gravitational energy, albeit in minuscule amounts. Since this is unlikely to be the case in nature, it seems that the quantum theory will have to modify not only Maxwell's electrodynamics but also the new theory of gravitation  \cite{Einstein1,Einstein2,Einstein:1917zz}. This remark raised two interconnecting questions. 

A possible modification of the electrodynamics and the general relativity follows from quantum-mechanical principles that govern the interaction of elementary particles \cite{Dirac:1927dy,Fermi:1932xva, Fermi1954}.  The answer to it was found  when the quantum correction to the classical action of electrodynamics was discovered by Heisenberg and Euler \cite{Heisenberg:1936nmg,Schwinger:1951nm} and in the corresponding  calculation of quantum correction to the Hilbert action in general relativity by DeWitt, 't Hooft, Veltman and other authors  \cite{DeWitt:1967uc, tHooft:1974toh, Goroff:1985th,  Bunch:1978yq}. The quantum-mechanical corrections to the classical action of the Yang-Mills theory have been also found in \cite{Savvidy:1977as,Matinyan:1976mp, Batalin:1976uv, Savvidy:2019grj}.   The string theory as well  provided the gravitational effective actions by calculating the background beta-function \cite{Green:1987sp, Green:1987mn}.       The second part of the Einstein's  remark concerning the classical gravitational instability of the atoms seems to be addressed to the theory that unifies the quantum mechanics and the general relativity.  

{\it Has a modern development of the quantum theory of gravity an answer to these questions  and does there exist an elementary particle that mediates the gravitational interaction - the  graviton? } 

The string theory that unifies the quantum mechanics, gauge fields and the general relativity aims to provide answers to these challenging questions  \cite{Dyson:2013hbl, Tobar:2023ksi, Rothman:2006fp, Boughn:2006st, Marletto:2017kzi, Carney:2024dsj, Manikandan:2025qgv}.  In the absence of fully satisfactory unification of quantum-mechanical principles and the general relativity the problem of gravitational stability of the atoms raised by Einstein is an interesting and challenging problem.  The stability problem was   discussed in     \cite{Blim}, and the answer should be within the realms of the quantum mechanics \cite{Dirac:1927dy,Fermi:1932xva, Fermi1954}.  The stability of the $\vert 1s \rangle$ state can be  demonstrated  by showing that  the quantum-mechanical quadrupole  matrix elements vanish (see discussion in  the concluding section).

In the classical electrodynamics the problem of atoms stability was raised by Lorentz  and in the classical general relativity by Einstein \cite{Einstein1,Einstein2}. Einstein raised the question of the consistency of the general relativity in its existing  form  due to the apparent conflict with the stability of atomic systems through the radiation of gravitational waves. 

 The quantum mechanics ensures  the electromagnetic stability of the atoms, while in the absence of fully satisfactory unification of the quantum-mechanical principles and the general relativity the problem of the gravitational stability of the atoms raised by Einstein is an important  and challenging problem. The string theory that unifies the gauge field theories and gravity aims to provide a consistent solution of gravitational stability of atoms. Here we consider a possible solution of the gravitational stability of the hydrogen atom by demonstrating that the transition rate vanishes due to the cancellation of the quantum-mechanical matrix elements of the quadrupole momentum of the neutral hydrogen atoms in its $\vert 1s \rangle$  state.

\noindent
{\it  Electromagnetic stability of the hydrogen $\vert 1s \rangle$ state.}
  In classical electrodynamics any accelerating charge radiates energy:
\be
\bar{\frac{d \CE}{d t}} = \frac{2 e^2 }{3 c^3}~ \ddot{\vec{x}}^2 .\nn
\ee
 For the circular motion of an electron  of mass $m$  and interacting with a proton through the Coulomb force    the dipole radiation would be:
$
\bar{\frac{d \CE}{d t}} = \frac{2 e^2~ a^{2}  \omega^4 }{3 c^3}, \nn
$
where   $a=  \hbar^2/m e^2$ is the Bohr radius,  $\omega^2 = e^2 /m a^3$, and the energy radiating per unit time is
\be
\bar{\frac{d \CE}{d t}} = \frac{4}{3} \Big(\frac{e^2}{\hbar c}\Big)^5  \Big(\frac{m e^4}{2 \hbar^2}\Big) \Big(\frac{m c^2}{ \hbar}\Big),\nn
\ee
where $Ryd= m e^4/2 \hbar^2  $ is the Rydberg constant  and $\alpha = e^2/ \hbar c $ is the  fine-structure constant.   If the electron is assumed to orbit in a circle and radiates energy continuously, the electron would rapidly spiral into the nucleus with a fall time of
\be
\frac{1}{T} =\frac{\bar{\frac{d \CE}{d t}}}{Ryd}=  \frac{4}{3} \Big(\frac{e^2}{\hbar c}\Big)^5   \Big(\frac{m c^2}{ \hbar}\Big).     \nn
\ee
The numerical value is $T   \approx 4.66 \times 10^{-11} sec$ and the atoms would instantly collapse.  In the quantum mechanics the transition rate of the spontaneous dipole  radiation of the electron from state $m$ to  state $n$  is given in (\ref{fermiabs}).
For the hydrogen atom the matrix element $ \langle n  \vert \vec{x} \vert  m \rangle$ for states  $\vert  n \rangle = \vert  m \rangle = \vert 1s \rangle$  vanishes:
$
\langle 1s  \vert \vec{x} \vert  1s \rangle = \int R^2_{10} Y^2_{0,0}(\theta, \phi)   r^2  d r \sin \theta d\theta d \phi =0.
$
The  $\vert  1s \rangle $  state is stable  and realises the  ground state of the hydrogen atom. 

\noindent
{\it  Gravitational stability of the hydrogen $\vert 1s \rangle$ state.}  A similar consideration  can be applied  to the gravitational radiation. The energy density of the classical quadrupole gravitational radiation  has the following form \cite{Einstein2,Landau1975}:
\be 
\bar{\frac{d \CE}{d t}} = \frac{G}{45 c^5} \dddot{D}_{ij}\dddot{D}_{ij}, \nn
\ee
where the quadrupole momentum tensor  is
$
D_{ij}= \int \rho (3 x_i x_j -\delta_{ij} \vec{x}^2) d V \nn
$
and $\rho$ is the density of mass.   For the circular motion of the electron  of mass $m$ interacting with a  proton through the Coulomb force  one can get:
$
\bar{\frac{d \CE}{d t}} = \frac{32 G~ m^{2} a^4 \omega^6 }{5 c^5},\nn
$
where   $a=  \hbar^2/m e^2$ and $\omega^2 = e^2 /m a^3$.  The intensity of radiation energy  is
\be
\bar{\frac{d \CE}{d t}}   
=\frac{64}{5}  \frac{ G m^2}{ \hbar c}   \Big( \frac{ e^2}{\hbar c}\Big)^6   \Big(\frac{ m e^4}{2 \hbar^2}\Big) \Big(\frac{ \ m c^2  }{\hbar} \Big)  \nn
\ee
and is of order $ 5.74 \times 10^{-47} ~\frac{erg}{sec}$. The electron orbiting a proton will continuously radiate gravitational waves and would spiral into the nucleus with a fall time
\be
\frac{1}{T}= { \bar{\frac{d \CE}{d t}} \over  Ryd }  =  \frac{64}{5}   {  G   m^2 \over \hbar c }  \Big( \frac{ e^2}{\hbar c}\Big)^{6}  \Big(\frac{ \ m c^2  }{\hbar} \Big)   \nn
\ee
of order $T  \approx 3.8 \times 10^{35} sec $. This time is larger than the Hubble time, but still is nonzero. 
  The quantum-mechanical  quadrupole  transition rate is (\ref{QMEinstein2}):  
 \be  
(Transition~ Rate~  m \rightarrow n)_g =  \frac{2 G \omega^5_{mn}}{45 \hbar c^5} ~ \vert \langle n \vert  D_{ij} \vert m \rangle  \vert^2\nn
\ee
and the matrix elements  for the  states $\vert  n \rangle = \vert  m \rangle = \vert 1s \rangle$    vanish:
 \beqa
   \langle 1s \vert D_{ij} \vert 1s \rangle =m   \int  R^2_{10} Y^2_{0,0} ~  (3 x_{i} x_{i} - \delta_{ij} r^2)     r^2 dr  \sin \theta d\theta d \phi=0,~~~~~~i,j=1,2,3,\nn
 \eeqa
and the  state $\vert  1s \rangle $  of the hydrogen atom  is stable against the gravitational radiation.




\end{appendices}

\bibliography{gravityatom}


\begin{thebibliography}{54}
\ifx \bisbn   \undefined \def \bisbn  #1{ISBN #1}\fi
\ifx \binits  \undefined \def \binits#1{#1}\fi
\ifx \bauthor  \undefined \def \bauthor#1{#1}\fi
\ifx \batitle  \undefined \def \batitle#1{#1}\fi
\ifx \bjtitle  \undefined \def \bjtitle#1{#1}\fi
\ifx \bvolume  \undefined \def \bvolume#1{\textbf{#1}}\fi
\ifx \byear  \undefined \def \byear#1{#1}\fi
\ifx \bissue  \undefined \def \bissue#1{#1}\fi
\ifx \bfpage  \undefined \def \bfpage#1{#1}\fi
\ifx \blpage  \undefined \def \blpage #1{#1}\fi
\ifx \burl  \undefined \def \burl#1{\textsf{#1}}\fi
\ifx \doiurl  \undefined \def \doiurl#1{\url{https://doi.org/#1}}\fi
\ifx \betal  \undefined \def \betal{\textit{et al.}}\fi
\ifx \binstitute  \undefined \def \binstitute#1{#1}\fi
\ifx \binstitutionaled  \undefined \def \binstitutionaled#1{#1}\fi
\ifx \bctitle  \undefined \def \bctitle#1{#1}\fi
\ifx \beditor  \undefined \def \beditor#1{#1}\fi
\ifx \bpublisher  \undefined \def \bpublisher#1{#1}\fi
\ifx \bbtitle  \undefined \def \bbtitle#1{#1}\fi
\ifx \bedition  \undefined \def \bedition#1{#1}\fi
\ifx \bseriesno  \undefined \def \bseriesno#1{#1}\fi
\ifx \blocation  \undefined \def \blocation#1{#1}\fi
\ifx \bsertitle  \undefined \def \bsertitle#1{#1}\fi
\ifx \bsnm \undefined \def \bsnm#1{#1}\fi
\ifx \bsuffix \undefined \def \bsuffix#1{#1}\fi
\ifx \bparticle \undefined \def \bparticle#1{#1}\fi
\ifx \barticle \undefined \def \barticle#1{#1}\fi
\bibcommenthead
\ifx \bconfdate \undefined \def \bconfdate #1{#1}\fi
\ifx \botherref \undefined \def \botherref #1{#1}\fi
\ifx \url \undefined \def \url#1{\textsf{#1}}\fi
\ifx \bchapter \undefined \def \bchapter#1{#1}\fi
\ifx \bbook \undefined \def \bbook#1{#1}\fi
\ifx \bcomment \undefined \def \bcomment#1{#1}\fi
\ifx \oauthor \undefined \def \oauthor#1{#1}\fi
\ifx \citeauthoryear \undefined \def \citeauthoryear#1{#1}\fi
\ifx \endbibitem  \undefined \def \endbibitem {}\fi
\ifx \bconflocation  \undefined \def \bconflocation#1{#1}\fi
\ifx \arxivurl  \undefined \def \arxivurl#1{\textsf{#1}}\fi
\csname PreBibitemsHook\endcsname

\bibitem[\protect\citeauthoryear{Weinberg}{1964a}]{Weinberg:1964kqu}
\begin{barticle}
\bauthor{\bsnm{Weinberg}, \binits{S.}}:
\batitle{{Derivation of gauge invariance and the equivalence principle from
  Lorentz invariance of the S- matrix}}.
\bjtitle{Phys. Lett.}
\bvolume{9}(\bissue{4}),
\bfpage{357}--\blpage{359}
(\byear{1964})
\doiurl{10.1016/0031-9163(64)90396-8}
\end{barticle}
\endbibitem

\bibitem[\protect\citeauthoryear{Weinberg}{1964b}]{Weinberg:1964ew}
\begin{barticle}
\bauthor{\bsnm{Weinberg}, \binits{S.}}:
\batitle{{Photons and Gravitons in $S$-Matrix Theory: Derivation of Charge
  Conservation and Equality of Gravitational and Inertial Mass}}.
\bjtitle{Phys. Rev.}
\bvolume{135},
\bfpage{1049}--\blpage{1056}
(\byear{1964})
\doiurl{10.1103/PhysRev.135.B1049}
\end{barticle}
\endbibitem

\bibitem[\protect\citeauthoryear{Weinberg}{1965a}]{Weinberg:1965nx}
\begin{barticle}
\bauthor{\bsnm{Weinberg}, \binits{S.}}:
\batitle{{Infrared photons and gravitons}}.
\bjtitle{Phys. Rev.}
\bvolume{140},
\bfpage{516}--\blpage{524}
(\byear{1965})
\doiurl{10.1103/PhysRev.140.B516}
\end{barticle}
\endbibitem

\bibitem[\protect\citeauthoryear{Weinberg}{1965b}]{Weinberg:1965rz}
\begin{barticle}
\bauthor{\bsnm{Weinberg}, \binits{S.}}:
\batitle{{Photons and gravitons in perturbation theory: Derivation of Maxwell's
  and Einstein's equations}}.
\bjtitle{Phys. Rev.}
\bvolume{138},
\bfpage{988}--\blpage{1002}
(\byear{1965})
\doiurl{10.1103/PhysRev.138.B988}
\end{barticle}
\endbibitem

\bibitem[\protect\citeauthoryear{Gould}{1985}]{Gould}
\begin{barticle}
\bauthor{\bsnm{Gould}, \binits{R.J.}}:
\batitle{{The graviton luminosity of the sun and other stars}}.
\bjtitle{The Astrophysical Journal}
\bvolume{288},
\bfpage{789}--\blpage{794}
(\byear{1985})
\doiurl{10.1086/162848}
\end{barticle}
\endbibitem

\bibitem[\protect\citeauthoryear{Garc{\'\i}a-Cely and
  Ringwald}{2025}]{Garcia-Cely:2024ujr}
\begin{barticle}
\bauthor{\bsnm{Garc{\'\i}a-Cely}, \binits{C.}},
\bauthor{\bsnm{Ringwald}, \binits{A.}}:
\batitle{{Complete Gravitational-Wave Spectrum of the Sun}}.
\bjtitle{Phys. Rev. Lett.}
\bvolume{135}(\bissue{6}),
\bfpage{061001}
(\byear{2025})
\doiurl{10.1103/gtwg-pr41}
{\href{https://arxiv.org/abs/2407.18297}{{arXiv:2407.18297}}}
{[hep-ph]}
\end{barticle}
\endbibitem

\bibitem[\protect\citeauthoryear{Dyson and Williams}{2020}]{Dyson2020}
\begin{bbook}
\bauthor{\bsnm{Dyson}, \binits{J.E.}},
\bauthor{\bsnm{Williams}, \binits{D.A.}}:
\bbtitle{{The Physics of the Interstellar Medium }}.
\bpublisher{CRC Press},
\blocation{New York}
(\byear{2020}).
\doiurl{10.1201/9781003025030}
\end{bbook}
\endbibitem

\bibitem[\protect\citeauthoryear{{Peebles}}{1969}]{Peebles1969ApJ}
\begin{barticle}
\bauthor{\bsnm{{Peebles}}, \binits{P.J.E.}}:
\batitle{{Intergalactic Hydrogen}}.
\bjtitle{Astrophysical Journal}
\bvolume{157},
\bfpage{45}
(\byear{1969})
\doiurl{10.1086/150048}
\end{barticle}
\endbibitem

\bibitem[\protect\citeauthoryear{Ghiglieri and Laine}{2015}]{Ghiglieri_2015}
\begin{barticle}
\bauthor{\bsnm{Ghiglieri}, \binits{J.}},
\bauthor{\bsnm{Laine}, \binits{M.}}:
\batitle{Gravitational wave background from standard model physics: qualitative
  features}.
\bjtitle{Journal of Cosmology and Astroparticle Physics}
\bvolume{2015}(\bissue{07}),
\bfpage{022}--\blpage{022}
(\byear{2015})
\doiurl{10.1088/1475-7516/2015/07/022}
\end{barticle}
\endbibitem

\bibitem[\protect\citeauthoryear{Ringwald et~al.}{2021}]{Ringwald_2021}
\begin{barticle}
\bauthor{\bsnm{Ringwald}, \binits{A.}},
\bauthor{\bsnm{Schütte-Engel}, \binits{J.}},
\bauthor{\bsnm{Tamarit}, \binits{C.}}:
\batitle{Gravitational waves as a big bang thermometer}.
\bjtitle{Journal of Cosmology and Astroparticle Physics}
\bvolume{2021}(\bissue{03}),
\bfpage{054}
(\byear{2021})
\doiurl{10.1088/1475-7516/2021/03/054}
\end{barticle}
\endbibitem

\bibitem[\protect\citeauthoryear{Grishchuk and Sidorov}{1989}]{Grishchuk_1989}
\begin{barticle}
\bauthor{\bsnm{Grishchuk}, \binits{L.P.}},
\bauthor{\bsnm{Sidorov}, \binits{Y.V.}}:
\batitle{{On the quantum state of relic gravitons}}.
\bjtitle{{Classical and Quantum Gravity}}
\bvolume{6},
\bfpage{161}
(\byear{1989})
\doiurl{10.1088/0264-9381/6/9/002}
\end{barticle}
\endbibitem

\bibitem[\protect\citeauthoryear{Arvanitaki and
  Geraci}{2013}]{PhysRevLett.110.071105}
\begin{barticle}
\bauthor{\bsnm{Arvanitaki}, \binits{A.}},
\bauthor{\bsnm{Geraci}, \binits{A.A.}}:
\batitle{Detecting high-frequency gravitational waves with optically levitated
  sensors}.
\bjtitle{Phys. Rev. Lett.}
\bvolume{110},
\bfpage{071105}
(\byear{2013})
\doiurl{10.1103/PhysRevLett.110.071105}
\end{barticle}
\endbibitem

\bibitem[\protect\citeauthoryear{Dyson}{2013}]{Dyson:2013hbl}
\begin{barticle}
\bauthor{\bsnm{Dyson}, \binits{F.}}:
\batitle{{Is a graviton detectable?}}
\bjtitle{Int. J. Mod. Phys. A}
\bvolume{28},
\bfpage{1330041}
(\byear{2013})
\doiurl{10.1142/S0217751X1330041X}
\end{barticle}
\endbibitem

\bibitem[\protect\citeauthoryear{Tobar et~al.}{2024}]{Tobar:2023ksi}
\begin{barticle}
\bauthor{\bsnm{Tobar}, \binits{G.}},
\bauthor{\bsnm{Manikandan}, \binits{S.K.}},
\bauthor{\bsnm{Beitel}, \binits{T.}},
\bauthor{\bsnm{Pikovski}, \binits{I.}}:
\batitle{{Detecting single gravitons with quantum sensing}}.
\bjtitle{Nature Commun.}
\bvolume{15}(\bissue{1}),
\bfpage{7229}
(\byear{2024})
\doiurl{10.1038/s41467-024-51420-8}
{\href{https://arxiv.org/abs/2308.15440}{{arXiv:2308.15440}}}
{[quant-ph]}
\end{barticle}
\endbibitem

\bibitem[\protect\citeauthoryear{Rothman and Boughn}{2006}]{Rothman:2006fp}
\begin{barticle}
\bauthor{\bsnm{Rothman}, \binits{T.}},
\bauthor{\bsnm{Boughn}, \binits{S.}}:
\batitle{{Can gravitons be detected?}}
\bjtitle{Found. Phys.}
\bvolume{36},
\bfpage{1801}--\blpage{1825}
(\byear{2006})
\doiurl{10.1007/s10701-006-9081-9}
{\href{https://arxiv.org/abs/gr-qc/0601043}{{arXiv:gr-qc/0601043}}}
\end{barticle}
\endbibitem

\bibitem[\protect\citeauthoryear{Boughn and Rothman}{2006}]{Boughn:2006st}
\begin{barticle}
\bauthor{\bsnm{Boughn}, \binits{S.}},
\bauthor{\bsnm{Rothman}, \binits{T.}}:
\batitle{{Aspects of graviton detection: Graviton emission and absorption by
  atomic hydrogen}}.
\bjtitle{Class. Quant. Grav.}
\bvolume{23},
\bfpage{5839}--\blpage{5852}
(\byear{2006})
\doiurl{10.1088/0264-9381/23/20/006}
{\href{https://arxiv.org/abs/gr-qc/0605052}{{arXiv:gr-qc/0605052}}}
\end{barticle}
\endbibitem

\bibitem[\protect\citeauthoryear{Marletto and Vedral}{2017}]{Marletto:2017kzi}
\begin{barticle}
\bauthor{\bsnm{Marletto}, \binits{C.}},
\bauthor{\bsnm{Vedral}, \binits{V.}}:
\batitle{{Gravitationally-induced entanglement between two massive particles is
  sufficient evidence of quantum effects in gravity}}.
\bjtitle{Phys. Rev. Lett.}
\bvolume{119}(\bissue{24}),
\bfpage{240402}
(\byear{2017})
\doiurl{10.1103/PhysRevLett.119.240402}
{\href{https://arxiv.org/abs/1707.06036}{{arXiv:1707.06036}}}
{[quant-ph]}
\end{barticle}
\endbibitem

\bibitem[\protect\citeauthoryear{Carney}{2024}]{Carney:2024dsj}
\begin{bchapter}
\bauthor{\bsnm{Carney}, \binits{D.}}:
\bctitle{{Comments on Graviton Detection}}.
In: \bbtitle{{Gravity, Strings and Fields}: {A Conference in Honour of Gordon
  Semenoff}}
(\byear{2024}).
\doiurl{10.1007/978-3-031-91266-5_2}
\end{bchapter}
\endbibitem

\bibitem[\protect\citeauthoryear{Manikandan and
  Wilczek}{2025}]{Manikandan:2025qgv}
\begin{barticle}
\bauthor{\bsnm{Manikandan}, \binits{S.K.}},
\bauthor{\bsnm{Wilczek}, \binits{F.}}:
\batitle{{Complementary probes of gravitational radiation states}}.
\bjtitle{Phys. Rev. A}
\bvolume{112}(\bissue{4}),
\bfpage{043716}
(\byear{2025})
\doiurl{10.1103/83tt-tt57}
{\href{https://arxiv.org/abs/2505.11422}{{arXiv:2505.11422}}}
{[gr-qc]}
\end{barticle}
\endbibitem

\bibitem[\protect\citeauthoryear{Savvidy}{2025}]{Savvidy:2025rqt}
\begin{barticle}
\bauthor{\bsnm{Savvidy}, \binits{G.}}:
\batitle{{Schwinger{\textquoteright}s non-commutative coordinates for photons
  and gravitons. Duality between helicity and Dirac quantization conditions}}.
\bjtitle{J. Math. Phys.}
\bvolume{66}(\bissue{8}),
\bfpage{082302}
(\byear{2025})
\doiurl{10.1063/5.0279941}
\end{barticle}
\endbibitem

\bibitem[\protect\citeauthoryear{Einstein}{1918}]{Einstein2}
\begin{botherref}
\oauthor{\bsnm{Einstein}, \binits{A.}}:
Über gravitationswellen.
Sitzungsberichte der Königlich Preussischen Akademie der Wissenschaften.,
--154167
(1918)
\end{botherref}
\endbibitem

\bibitem[\protect\citeauthoryear{Landau and Lifschitz}{1975}]{Landau1975}
\begin{bbook}
\bauthor{\bsnm{Landau}, \binits{L.D.}},
\bauthor{\bsnm{Lifschitz}, \binits{E.M.}}:
\bbtitle{{vol.2: The Classical Theory of Fields (Fourth Edition) }}.
\bsertitle{Course of Theoretical Physics, Publisher Pergamon},
(\byear{1975})
\end{bbook}
\endbibitem

\bibitem[\protect\citeauthoryear{Weinberg}{1972}]{Weinberg:1972kfs}
\begin{bbook}
\bauthor{\bsnm{Weinberg}, \binits{S.}}:
\bbtitle{{Gravitation and Cosmology}: {Principles and Applications of the
  General Theory of Relativity}}.
\bpublisher{John Wiley and Sons},
\blocation{New York}
(\byear{1972})
\end{bbook}
\endbibitem

\bibitem[\protect\citeauthoryear{Fischer}{1994}]{Fischer_1994}
\begin{barticle}
\bauthor{\bsnm{Fischer}, \binits{U.}}:
\batitle{Transition probabilities for a rydberg atom in the field of a
  gravitational wave}.
\bjtitle{Classical and Quantum Gravity}
\bvolume{11},
\bfpage{463}
(\byear{1994})
\doiurl{10.1088/0264-9381/11/2/018}
\end{barticle}
\endbibitem

\bibitem[\protect\citeauthoryear{{Bunker} and
  et~al.}{2023}]{2023A&A...677A..88B}
\begin{barticle}
\bauthor{\bsnm{{Bunker}}, \binits{A.J.}},
\bauthor{\bsnm{al.}}:
\batitle{{JADES NIRSpec Spectroscopy of GN-z11: Lyman-{\ensuremath{\alpha}}
  emission and possible enhanced nitrogen abundance in a z = 10.60 luminous
  galaxy}}.
\bjtitle{Astronomy and Astrophysics}
\bvolume{677},
\bfpage{88}
(\byear{2023})
\doiurl{10.1051/0004-6361/202346159}
\end{barticle}
\endbibitem

\bibitem[\protect\citeauthoryear{{Heintz} and
  et~al.}{2025}]{2025A&A...693A..60H}
\begin{barticle}
\bauthor{\bsnm{{Heintz}}, \binits{K.E.}},
\bauthor{\bsnm{al.}}:
\batitle{{The JWST-PRIMAL archival survey: A JWST/NIRSpec reference sample for
  the physical properties and Lyman-{\ensuremath{\alpha}} absorption and
  emission of {\ensuremath{\sim}}600 galaxies at z = 5.0 {\ensuremath{-}}
  13.4}}.
\bjtitle{Astronomy and Astrophysics}
\bvolume{693},
\bfpage{60}
(\byear{2025})
\doiurl{10.1051/0004-6361/202450243}
{[astro-ph.GA]}
\end{barticle}
\endbibitem

\bibitem[\protect\citeauthoryear{Kashino et~al.}{2023}]{Kashino_2023}
\begin{barticle}
\bauthor{\bsnm{Kashino}, \binits{D.}},
\bauthor{\bsnm{Lilly}, \binits{S.J.}},
\bauthor{\bsnm{Matthee}, \binits{J.}},
\bauthor{\bsnm{Eilers}, \binits{A.-C.}},
\bauthor{\bsnm{Mackenzie}, \binits{R.}},
\bauthor{\bsnm{Bordoloi}, \binits{R.}},
\bauthor{\bsnm{Simcoe}, \binits{R.A.}}:
\batitle{Eiger: A large sample of [o iii]-emitting galaxies at 5.3
  {\ensuremath{-}} 6.9 and direct evidence for local reionization by galaxies}.
\bjtitle{Astrophysical Journal}
\bvolume{950}(\bissue{1}),
\bfpage{66}
(\byear{2023})
\doiurl{10.3847/1538-4357/acc588}
\end{barticle}
\endbibitem

\bibitem[\protect\citeauthoryear{Heintz et~al.}{2024}]{heintz2023}
\begin{barticle}
\bauthor{\bsnm{Heintz}, \binits{K.E.}},
\bauthor{\bsnm{Watson}, \binits{D.}},
\bauthor{\bsnm{Brammer}, \binits{G.}},
\bauthor{\bsnm{Vejlgaard}, \binits{S.}},
\bauthor{\bsnm{Hutter}, \binits{A.}},
\bauthor{\bsnm{Strait}, \binits{V.B.}},
\bauthor{\bsnm{Matthee}, \binits{J.}},
\bauthor{\bsnm{Oesch}, \binits{P.A.}},
\bauthor{\bsnm{Jakobsson}, \binits{P.}},
\bauthor{\bsnm{Tanvir}, \binits{N.R.}},
\bauthor{\bsnm{Laursen}, \binits{P.}},
\bauthor{\bsnm{Naidu}, \binits{R.P.}},
\bauthor{\bsnm{Mason}, \binits{C.A.}},
\bauthor{\bsnm{Killi}, \binits{M.}},
\bauthor{\bsnm{Jung}, \binits{I.}},
\bauthor{\bsnm{Hsiao}, \binits{T.Y.-Y.}},
\bauthor{\bsnm{Abdurro'uf}},
\bauthor{\bsnm{Coe}, \binits{D.}},
\bauthor{\bsnm{Haro}, \binits{P.A.}},
\bauthor{\bsnm{Finkelstein}, \binits{S.L.}},
\bauthor{\bsnm{Toft.}, \binits{S.}}:
\batitle{Strong damped lyman-$\alpha$ absorption in young star-forming galaxies
  at $z=9-11$}.
\bjtitle{Science}
\bvolume{384},
\bfpage{890}
(\byear{2024})
\doiurl{10.1126/science.adj0343}
\end{barticle}
\endbibitem

\bibitem[\protect\citeauthoryear{{Sanders} et~al.}{2024}]{2024ApJ...962...24S}
\begin{barticle}
\bauthor{\bsnm{{Sanders}}, \binits{R.L.}},
\bauthor{\bsnm{{Shapley}}, \binits{A.E.}},
\bauthor{\bsnm{{Topping}}, \binits{M.W.}},
\bauthor{\bsnm{{Reddy}}, \binits{N.A.}},
\bauthor{\bsnm{{Brammer}}, \binits{G.B.}}:
\batitle{Direct t$e$-based metallicities of z = 2 {\ensuremath{-}} 9 galaxies
  with jwst/nirspec: Empirical metallicity calibrations applicable from
  reionisation to cosmic noon}.
\bjtitle{Astrophysical Journal}
\bvolume{962}(\bissue{1}),
\bfpage{24}
(\byear{2024})
\doiurl{10.3847/1538-4357/ad15fc}
\end{barticle}
\endbibitem

\bibitem[\protect\citeauthoryear{Einstein}{1917}]{Einstein:1917zz}
\begin{barticle}
\bauthor{\bsnm{Einstein}, \binits{A.}}:
\batitle{{Zur Quantentheorie der Strahlung}}.
\bjtitle{Phys. Z.}
\bvolume{18},
\bfpage{121}--\blpage{128}
(\byear{1917})
\end{barticle}
\endbibitem

\bibitem[\protect\citeauthoryear{Dirac}{1927}]{Dirac:1927dy}
\begin{barticle}
\bauthor{\bsnm{Dirac}, \binits{P.A.M.}}:
\batitle{{Quantum theory of emission and absorption of radiation}}.
\bjtitle{Proc. Roy. Soc. Lond. A}
\bvolume{114},
\bfpage{243}
(\byear{1927})
\doiurl{10.1098/rspa.1927.0039}
\end{barticle}
\endbibitem

\bibitem[\protect\citeauthoryear{Fermi}{1932}]{Fermi:1932xva}
\begin{barticle}
\bauthor{\bsnm{Fermi}, \binits{E.}}:
\batitle{{Quantum Theory of Radiation}}.
\bjtitle{Rev. Mod. Phys.}
\bvolume{4}(\bissue{1}),
\bfpage{87}--\blpage{132}
(\byear{1932})
\doiurl{10.1103/RevModPhys.4.87}
\end{barticle}
\endbibitem

\bibitem[\protect\citeauthoryear{Fermi}{1960}]{Fermi1954}
\begin{bbook}
\bauthor{\bsnm{Fermi}, \binits{E.}}:
\bbtitle{{Notes on Quantum Mechanics}}.
\bpublisher{The University of Chicago Press},
\blocation{Chicago}
(\byear{1960})
\end{bbook}
\endbibitem

\bibitem[\protect\citeauthoryear{Kramida et~al.}{2026}]{Kramida}
\begin{barticle}
\bauthor{\bsnm{Kramida}, \binits{A.}},
\bauthor{\bsnm{Ralchenko}, \binits{Y.}},
\bauthor{\bsnm{Reader}, \binits{J.}},
\bauthor{\bsnm{al.}}:
\batitle{{NIST Atomic Spectra Database (version 5.12)}}.
\bjtitle{National Institute of Standards and Technology, Gaithersburg, MD.}
(\byear{2026})
\doiurl{10.18434/T4W30F}
\end{barticle}
\endbibitem

\bibitem[\protect\citeauthoryear{{Gueymard}}{2004}]{2004SoEn...76..423G}
\begin{barticle}
\bauthor{\bsnm{{Gueymard}}, \binits{C.A.}}:
\batitle{{The sun's total and spectral irradiance for solar energy applications
  and solar radiation models}}.
\bjtitle{Solar Energy}
\bvolume{76}(\bissue{4}),
\bfpage{423}--\blpage{453}
(\byear{2004})
\doiurl{10.1016/j.solener.2003.08.039}
\end{barticle}
\endbibitem

\bibitem[\protect\citeauthoryear{Delbouille et~al.}{1981}]{Delbouille}
\begin{bbook}
\bauthor{\bsnm{Delbouille}, \binits{L.}},
\bauthor{\bsnm{Roland}, \binits{G.}},
\bauthor{\bsnm{Brault}, \binits{J.}},
\bauthor{\bsnm{Testerman}, \binits{L.}}:
\bbtitle{{Photometric Atlas of the Solar Spectrum}},
(\byear{1981}).
\burl{http://bass2000.obspm.fr/solar spect.php}
\end{bbook}
\endbibitem

\bibitem[\protect\citeauthoryear{{Abrams} et~al.}{1996}]{1996ApOpt..35.2747A}
\begin{barticle}
\bauthor{\bsnm{{Abrams}}, \binits{M.C.}},
\bauthor{\bsnm{{Goldman}}, \binits{A.}},
\bauthor{\bsnm{{Gunson}}, \binits{M.R.}},
\bauthor{\bsnm{{Rinsland}}, \binits{C.P.}},
\bauthor{\bsnm{{Zander}}, \binits{R.}}:
\batitle{{Observations of the infrared solar spectrum from space by the ATMOS
  experiment}}.
\bjtitle{Applied Optics}
\bvolume{35}(\bissue{16}),
\bfpage{2747}--\blpage{2751}
(\byear{1996})
\doiurl{10.1364/AO.35.002747}
\end{barticle}
\endbibitem

\bibitem[\protect\citeauthoryear{Asplund et~al.}{2009}]{Asplund_2009}
\begin{barticle}
\bauthor{\bsnm{Asplund}, \binits{M.}},
\bauthor{\bsnm{Grevesse}, \binits{N.}},
\bauthor{\bsnm{Sauval}, \binits{A.J.}},
\bauthor{\bsnm{Scott}, \binits{P.}}:
\batitle{The chemical composition of the sun}.
\bjtitle{Annual Review of Astronomy and Astrophysics}
\bvolume{47}(\bissue{1}),
\bfpage{481}--\blpage{522}
(\byear{2009})
\doiurl{10.1146/annurev.astro.46.060407.145222}
\end{barticle}
\endbibitem

\bibitem[\protect\citeauthoryear{{Lemaire, P.} et~al.}{2012}]{refId0}
\begin{barticle}
\bauthor{\bsnm{{Lemaire, P.}}},
\bauthor{\bsnm{{Vial, J.-C.}}},
\bauthor{\bsnm{{Curdt, W.}}},
\bauthor{\bsnm{{Sch\"uhle, U.}}},
\bauthor{\bsnm{{Woods, T. N.}}}:
\batitle{The solar hydrogen lyman-alpha lyman-beta ratio}.
\bjtitle{A\&A}
\bvolume{542},
\bfpage{25}
(\byear{2012})
\doiurl{10.1051/0004-6361/201219026}
\end{barticle}
\endbibitem

\bibitem[\protect\citeauthoryear{Ning et~al.}{2026}]{ning2026}
\begin{botherref}
\oauthor{\bsnm{Ning}, \binits{Y.}},
\oauthor{\bsnm{Cai}, \binits{Z.}},
\oauthor{\bsnm{Jiang}, \binits{L.}},
\oauthor{\bsnm{Guo}, \binits{Y.}},
\oauthor{\bsnm{Li}, \binits{Q.}},
\oauthor{\bsnm{Yu}, \binits{S.-Y.}},
\oauthor{\bsnm{Yu}, \binits{X.}},
\oauthor{\bsnm{Zheng}, \binits{Z.-Y.}}:
An Updated Characterization of Luminous Lyman-alpha emitters at the End of
  Reionization
(2026).
\url{https://arxiv.org/abs/2605.14313}
\end{botherref}
\endbibitem

\bibitem[\protect\citeauthoryear{{Einstein}}{1916}]{Einstein1}
\begin{barticle}
\bauthor{\bsnm{{Einstein}}, \binits{A.}}:
\batitle{{N{\"a}herungsweise Integration der Feldgleichungen der Gravitation}}.
\bjtitle{Sitzungsberichte der Königlich Preussischen Akademie der
  Wissenschaften.}
\bvolume{1},
\bfpage{688}--\blpage{696}
(\byear{1916})
\end{barticle}
\endbibitem

\bibitem[\protect\citeauthoryear{Heisenberg and
  Euler}{1936}]{Heisenberg:1936nmg}
\begin{barticle}
\bauthor{\bsnm{Heisenberg}, \binits{W.}},
\bauthor{\bsnm{Euler}, \binits{H.}}:
\batitle{{Consequences of Dirac's theory of positrons}}.
\bjtitle{Z. Phys.}
\bvolume{98}(\bissue{11-12}),
\bfpage{714}--\blpage{732}
(\byear{1936})
\doiurl{10.1007/BF01343663}
{\href{https://arxiv.org/abs/physics/0605038}{{arXiv:physics/0605038}}}
\end{barticle}
\endbibitem

\bibitem[\protect\citeauthoryear{Schwinger}{1951}]{Schwinger:1951nm}
\begin{barticle}
\bauthor{\bsnm{Schwinger}, \binits{J.S.}}:
\batitle{{On gauge invariance and vacuum polarization}}.
\bjtitle{Phys. Rev.}
\bvolume{82},
\bfpage{664}--\blpage{679}
(\byear{1951})
\doiurl{10.1103/PhysRev.82.664}
\end{barticle}
\endbibitem

\bibitem[\protect\citeauthoryear{DeWitt}{1967}]{DeWitt:1967uc}
\begin{barticle}
\bauthor{\bsnm{DeWitt}, \binits{B.S.}}:
\batitle{{Quantum Theory of Gravity. 3. Applications of the Covariant Theory}}.
\bjtitle{Phys. Rev.}
\bvolume{162},
\bfpage{1239}--\blpage{1256}
(\byear{1967})
\doiurl{10.1103/PhysRev.162.1239}
\end{barticle}
\endbibitem

\bibitem[\protect\citeauthoryear{'t~Hooft and Veltman}{1974}]{tHooft:1974toh}
\begin{barticle}
\bauthor{\bsnm{Hooft}, \binits{G.}},
\bauthor{\bsnm{Veltman}, \binits{M.J.G.}}:
\batitle{{One-loop divergencies in the theory of gravitation}}.
\bjtitle{Ann. Inst. H. Poincare Phys. Theor. A}
\bvolume{20}(\bissue{1}),
\bfpage{69}--\blpage{94}
(\byear{1974})
\doiurl{10.1142/9789814539395_0001}
\end{barticle}
\endbibitem

\bibitem[\protect\citeauthoryear{Goroff and Sagnotti}{1986}]{Goroff:1985th}
\begin{barticle}
\bauthor{\bsnm{Goroff}, \binits{M.H.}},
\bauthor{\bsnm{Sagnotti}, \binits{A.}}:
\batitle{{The Ultraviolet Behavior of Einstein Gravity}}.
\bjtitle{Nucl. Phys. B}
\bvolume{266},
\bfpage{709}--\blpage{736}
(\byear{1986})
\doiurl{10.1016/0550-3213(86)90193-8}
\end{barticle}
\endbibitem

\bibitem[\protect\citeauthoryear{Bunch and Davies}{1978}]{Bunch:1978yq}
\begin{barticle}
\bauthor{\bsnm{Bunch}, \binits{T.S.}},
\bauthor{\bsnm{Davies}, \binits{P.C.W.}}:
\batitle{{Quantum Field Theory in de Sitter Space: Renormalization by Point
  Splitting}}.
\bjtitle{Proc. Roy. Soc. Lond. A}
\bvolume{360},
\bfpage{117}--\blpage{134}
(\byear{1978})
\doiurl{10.1098/rspa.1978.0060}
\end{barticle}
\endbibitem

\bibitem[\protect\citeauthoryear{Savvidy}{1977}]{Savvidy:1977as}
\begin{barticle}
\bauthor{\bsnm{Savvidy}, \binits{G.K.}}:
\batitle{{Infrared Instability of the Vacuum State of Gauge Theories and
  Asymptotic Freedom}}.
\bjtitle{Phys. Lett. B}
\bvolume{71},
\bfpage{133}--\blpage{134}
(\byear{1977})
\doiurl{10.1016/0370-2693(77)90759-6}
\end{barticle}
\endbibitem

\bibitem[\protect\citeauthoryear{Matinyan and Savvidy}{1978}]{Matinyan:1976mp}
\begin{barticle}
\bauthor{\bsnm{Matinyan}, \binits{S.G.}},
\bauthor{\bsnm{Savvidy}, \binits{G.K.}}:
\batitle{{Vacuum Polarization Induced by the Intense Gauge Field}}.
\bjtitle{Nucl. Phys. B}
\bvolume{134},
\bfpage{539}--\blpage{545}
(\byear{1978})
\doiurl{10.1016/0550-3213(78)90463-7}
\end{barticle}
\endbibitem

\bibitem[\protect\citeauthoryear{Batalin et~al.}{1977}]{Batalin:1976uv}
\begin{barticle}
\bauthor{\bsnm{Batalin}, \binits{I.A.}},
\bauthor{\bsnm{Matinyan}, \binits{S.G.}},
\bauthor{\bsnm{Savvidy}, \binits{G.K.}}:
\batitle{{Vacuum Polarization by a Source-Free Gauge Field}}.
\bjtitle{Sov. J. Nucl. Phys.}
\bvolume{26},
\bfpage{214}
(\byear{1977})
\end{barticle}
\endbibitem

\bibitem[\protect\citeauthoryear{Savvidy}{2020}]{Savvidy:2019grj}
\begin{barticle}
\bauthor{\bsnm{Savvidy}, \binits{G.}}:
\batitle{{From Heisenberg\textendash{}Euler Lagrangian to the discovery of
  Chromomagnetic Gluon Condensation}}.
\bjtitle{Eur. Phys. J. C}
\bvolume{80}(\bissue{2}),
\bfpage{165}
(\byear{2020})
\doiurl{10.1140/epjc/s10052-020-7711-6}
{\href{https://arxiv.org/abs/1910.00654}{{arXiv:1910.00654}}}
{[hep-th]}
\end{barticle}
\endbibitem

\bibitem[\protect\citeauthoryear{Green et~al.}{1988a}]{Green:1987sp}
\begin{bbook}
\bauthor{\bsnm{Green}, \binits{M.B.}},
\bauthor{\bsnm{Schwarz}, \binits{J.H.}},
\bauthor{\bsnm{Witten}, \binits{E.}}:
\bbtitle{{Superstring Theory. Vol. 1: Introduction}}.
\bsertitle{Cambridge Monographs on Mathematical Physics},
(\byear{1988})
\end{bbook}
\endbibitem

\bibitem[\protect\citeauthoryear{Green et~al.}{1988b}]{Green:1987mn}
\begin{bbook}
\bauthor{\bsnm{Green}, \binits{M.B.}},
\bauthor{\bsnm{Schwarz}, \binits{J.H.}},
\bauthor{\bsnm{Witten}, \binits{E.}}:
\bbtitle{{Superstring Theory. Vol. 2: Loop Amplitudes, Anomalies and
  Phenomenology}}.
\bsertitle{Cambridge Monographs on Mathematical Physics},
(\byear{1988})
\end{bbook}
\endbibitem

\bibitem[\protect\citeauthoryear{collection of the earliest contributions to
  quantum gravity~research. A.S.Blum and R.Dean}{2018}]{Blim}
\begin{botherref}
Quantum gravity in the first half of the twentieth century: A sourcebook.
In: \oauthor{\bsnm{A.S.Blum}, \binits{A.}},
\oauthor{\bsnm{R.Dean}} (eds.)
{Berlin: Max-Planck-Gesellschaft zur Förderung der Wissenschaften.}
(2018)
\end{botherref}
\endbibitem

\end{thebibliography}

\end{document}